\documentclass[aps,preprint,showkeys]{revtex4}

\usepackage{amsmath}
\usepackage{amsfonts}
\usepackage{amssymb}
\usepackage{amsthm}
\usepackage{mathrsfs}
\usepackage{graphicx}
\usepackage{color}
\usepackage{physics}
\usepackage{float}
\usepackage{booktabs}

\usepackage[colorlinks=true,linkcolor=blue,citecolor=blue]{hyperref}

\usepackage{placeins}

\begin{document}
\newcommand{\be}{\begin{equation}}
\newcommand{\ee}{\end{equation}}
\newcommand{\bea}{\begin{eqnarray}}
\newcommand{\eea}{\end{eqnarray}}

\title{Rotation-Induced Effective Anisotropy in White Dwarfs as a Newtonian Benchmark with Relativistic Scale Assessment}

\author{ Aray Muratkhan$^{1}$, Aliya Taukenova$^{1}$, Saken Toktarbay$^{1*}$ and Hernando Quevedo$^{1,2,3}$}

\affiliation{
$^1$ Department of Theoretical and Nuclear Physics, \\
Al-Farabi Kazakh National University, Almaty, 050040, Kazakhstan \\
$^2$ Instituto de Ciencias Nucleares, Universidad Nacional Aut\'onoma de M\'exico, AP 70543, M\'exico, 04510,CDMX, Mexico \\
$^3$ Dipartimento di Fisica and ICRA, Universit\`a di Roma ``La Sapienza", Piazzale Aldo Moro 5, Rome,00185,Lazio, Italy \\
$^{*}$Corresponding author: {saken.toktarbay@kaznu.edu.kz}
}


\begin{abstract}
We develop a one-dimensional Newtonian reduction for uniformly rotating cold white dwarfs
in which the angle-averaged centrifugal support is represented by an effective anisotropic
term. From the stationary Euler equation, using
\(\langle\sin^2\theta\rangle=2/3\), the rotational contribution becomes $ \Delta_{\rm rot}(r)=\frac{1}{3}\rho(r)\Omega^2 r^2 .
$
The mapping keeps the spin frequency explicit while preserving a one-dimensional
hydrostatic system. With the Chandrasekhar degenerate-electron equation of state, we
compute sequences over
\(\rho_c\in[10^6,10^{11}]~{\rm g\,cm^{-3}}\) for rotation proxies
\(f=\Omega/\Omega_{K,0}(\rho_c)\leq0.35\). The high-density readout shows monotonic
increases of mass and radius with \(f\), with a percent-level mass shift for the largest
retained proxy. Applicability is checked on the rotating configurations through
sub-Keplerian diagnostics and the bulk-interior measure
\(\mathcal{A}_{10^{-2}}\). We further compare the reduced rotational correction with an
auxiliary quasi-two-dimensional reconstruction and a static isotropic Tolman--Oppenheimer--Volkoff reference
sequence. These scale checks show that the reduced model remains useful for controlled
trend-level surveys in the slow-rotation regime, while rotational and static relativistic
corrections can both become percent-level effects at high central density. The construction
provides a transparent Newtonian benchmark for future axisymmetric and relativistic
rotating white-dwarf calculations.
\end{abstract}

\maketitle

\section{Introduction}
\label{sec:introduction}

White dwarfs (WDs) provide a well-defined laboratory for testing the interplay between
degenerate-matter microphysics, gravity, and rotation. Their basic mass--radius behaviour is
well described by the cold degenerate-electron equation of state and its refinements
\citep{Chandrasekhar1931,Chandrasekhar1939,HamadaSalpeter1961,Nauenberg1972,
ShapiroTeukolsky1983}. In many stellar-structure applications WDs are treated as slowly
rotating objects, and this is often justified for isolated field WDs whose spin periods are
typically measured on timescales of hours to days. Observationally, however, the distribution
is broad: pulsating WDs, magnetic WDs, accreting systems, and compact binaries include
objects with substantially shorter periods \citep{Kawaler2015,Hermes2017ApJS,Brinkworth2013ApJ,Ferrario2015SSRv,
Kilic2021ApJL,Warner1995CVs}. Rotation can therefore be a negligible correction in some parts of parameter
space and a measurable contribution to global quantities in others.

A self-consistent treatment of rotation is not one-dimensional. Even in Newtonian gravity,
uniform rotation breaks spherical symmetry, deforms the stellar surface, and introduces a
latitude-dependent force balance. Classical descriptions of rotating self-gravitating fluids
are based on slow-rotation expansions, virial and ellipsoidal arguments, or self-consistent
field methods for axisymmetric equilibria
\citep{Chandrasekhar1969,Tassoul1978,OstrikerBodenheimer1968,Hachisu1986ApJS,
KomatsuEriguchiHachisu1989,EriguchiMuller1985}. In relativistic stellar modelling, the
standard perturbative route begins with the Hartle--Thorne slow-rotation formalism
\citep{Hartle1967,HartleThorne1968}, while fully rotating configurations require genuinely
two-dimensional numerical treatments \citep{CookShapiroTeukolsky1992,Stergioulas2003,
FriedmanStergioulas2013}. For WDs, relativistic and rotational effects have also been studied
in the context of compact high-density configurations and uniformly rotating sequences
\citep{Boshkayev2013ApJ,Rotondo2011,Boshkayev2014}.

At the same time, reduced one-dimensional models remain useful when the aim is not to
construct an exact rotating equilibrium, but to survey broad central-density ranges and to
estimate the size of rotation-induced trends. Such a reduction must be used with care. It
should state which part of the rotational force is retained, which geometric information is
lost, and how the resulting configurations are checked for physical admissibility. This is the
motivation of the present work.

We consider cold, uniformly rotating WDs and keep the gravitational sector spherically
symmetric. Starting from the stationary Euler equation in a uniformly rotating frame, we
project the centrifugal acceleration onto the spherical radial direction and replace the angular
factor by its spherical average, \(\langle \sin^2\theta\rangle=2/3\). The resulting radial balance
contains an additional support term proportional to \(\rho\Omega^2 r\). By comparing this
angle-averaged equation with the Newtonian anisotropic hydrostatic equation
\citep{BowersLiang1974,HerreraSantos1997,DevGleiser2002}, the rotational
contribution can be written as
\begin{equation}
\Delta_{\rm rot}(r)=\frac{1}{3}\rho(r)\Omega^2 r^2 .
\label{eq:DeltaRot}
\end{equation}
Here \(\Delta_{\rm rot}\) is not a microscopic pressure anisotropy of the stellar matter. It is an
effective one-dimensional representation of the angle-averaged centrifugal support. The
model therefore does not reproduce the local two-dimensional equilibrium of a rotating WD,
but it keeps the spin frequency explicit and gives a controlled way to follow rotational
trends in \(M(\rho_c)\), \(R(\rho_c)\), and related diagnostic quantities.

The structure equations are closed with the Chandrasekhar cold degenerate-electron
equation of state. We compute sequences over
\[
\rho_c\in[10^6,10^{11}]~{\rm g\,cm^{-3}},
\]
and parameterize rotation by
\[
f\equiv \Omega/\Omega_{K,0}(\rho_c),
\]
where \(\Omega_{K,0}\) is evaluated from the non-rotating Newtonian reference model at the
same central density. Since this proxy is defined with respect to the baseline configuration,
it is not itself a physical bound. Each rotating solution is therefore checked using
diagnostics evaluated on the rotating configuration,
\begin{equation}
\Omega_K(M,R)=\sqrt{\frac{GM}{R^3}},
\qquad
\epsilon \equiv \frac{\Omega^2 R^3}{GM}.
\label{eq:KeplerEps}
\end{equation}
We require \(\Omega/\Omega_K<1\) as a consistency condition against mass shedding. We also
monitor the relative size of the mapped term in the bulk interior through
\begin{equation}
\mathcal{A}_{10^{-2}}\equiv
\max_{p_r/p_c\ge 10^{-2}}
\left(\frac{\Delta_{\rm rot}}{p_r}\right),
\label{eq:A10m2}
\end{equation}
which excludes the low-pressure surface layer where \(\Delta_{\rm rot}/p_r\) is dominated by
boundary behaviour. Together with a force-balance diagnostic, this defines a
multi-criterion applicability envelope for the rotation proxy.

The reduced model is deliberately limited in scope. The replacement
\(\sin^2\theta\rightarrow\langle\sin^2\theta\rangle\) removes the latitude dependence of the
centrifugal force, and therefore cannot describe the equatorial bulge, polar flattening,
oblateness, or the true mass-shedding surface. These quantities require self-consistent
axisymmetric equilibria. In this paper the reduction is used only as a monitored
slow-to-moderate rotation benchmark.

We also include a static Tolman--Oppenheimer--Volkoff (TOV) scale assessment
\citep{Tolman1939,OppenheimerVolkoff1939}. This is not a relativistic extension of the
rotating model. Instead, we compare the size of the reduced Newtonian rotational correction
with the static relativistic correction obtained from a non-rotating isotropic TOV reference
sequence computed with the same pressure--density relation. The purpose is to identify the
density range in which rotational and static relativistic effects can both contribute at the
percent level to the same global observables. This comparison clarifies where the present
Newtonian benchmark remains adequate as a trend-level tool and where relativistic
slow-rotation or fully rotating models become necessary.

The paper is organized as follows. Section~\ref{sec:baseline} introduces the non-rotating
Newtonian reference sequence and the Chandrasekhar equation of state. Section~III derives
the rotation--anisotropy mapping and describes the numerical implementation. Section~IV
defines the applicability diagnostics and the reference axisymmetric formulation. Section~V
presents the reduced-model results, including the global sequences, the slow-rotation
envelope, and the quasi-two-dimensional consistency check. Section~VI gives the static TOV
scale assessment. Section~\ref{sec:disc_conc} summarizes the conclusions and limitations.


\section{Non-rotating isotropic Newtonian white dwarfs}
\label{sec:baseline}

We begin with a reference sequence of non-rotating, isotropic Newtonian white dwarfs supported by cold electron
degeneracy pressure at fixed composition \citep{Chandrasekhar1931,Chandrasekhar1939,ShapiroTeukolsky1983}. This baseline provides the benchmark families $M(\rho_c)$ and $R(\rho_c)$,
and the corresponding mass--radius relation $M(R)$, against which the rotation-induced models are compared
\citep{HamadaSalpeter1961}.
Throughout we assume a constant mean molecular weight per electron $\mu_e$ (e.g. $\mu_e\simeq 2$ for CO compositions)
and a barotropic equation of state $p=p(\rho)$ \citep{Chandrasekhar1939,ShapiroTeukolsky1983}.

\subsection{Structure equations}
\label{subsec:baseline_eqs}

Under spherical symmetry and hydrostatic equilibrium, the stellar structure is governed by the equilibrium equations
\begin{align}
\frac{\dd p}{\dd r} &= - \frac{G\, m(r)\, \rho(r)}{r^2},
\label{eq:hse_newton_base}\\
\frac{\dd m}{\dd r} &= 4\pi r^2 \rho(r),
\label{eq:mass_cont_base}
\end{align}
with conditions $m(0)=0$ and $\rho(0)=\rho_c$ at the center of the source. The stellar surface is defined by
\begin{equation}
p(R)=0,\qquad M \equiv m(R).
\label{eq:surface_base}
\end{equation}
Equations \eqref{eq:hse_newton_base}--\eqref{eq:surface_base} are closed by the Chandrasekhar degenerate-electron
equation of state summarized below \citep{Chandrasekhar1939,ShapiroTeukolsky1983}.

\subsection{Chandrasekhar equation of state}
\label{subsec:baseline_eos}

We use the standard parametric form in terms of the relativity parameter
\begin{equation}
x \equiv \frac{p_F}{m_e c},
\end{equation}
where $m_e$ and $p_F$ are the electron mass and the electron Fermi momentum, respectively. The pressure and density are then written as
\begin{align}
p(x) &= K\left[x(2x^2-3)\sqrt{1+x^2}+3\sinh^{-1}(x)\right],
\label{eq:chandra_p_base}\\
\rho(x) &= \rho_0\,x^3,
\label{eq:chandra_rho_base}
\end{align}
with constants
\begin{equation}
K=\frac{\pi m_e^4 c^5}{3h^3},\qquad
\rho_0=\mu_e m_u\,\frac{8\pi (m_e c)^3}{3h^3}
\label{eq:K_rho0_base}
\end{equation}
\citep{Chandrasekhar1939,ShapiroTeukolsky1983}.
Thus the barotrope $p(\rho)$ is obtained via $x=(\rho/\rho_0)^{1/3}$ \citep{Chandrasekhar1939}.

\subsection{Dimensionless form}
\label{subsec:baseline_dimless}

This particular representation of the structure equations is obtained by using the notation 
\begin{equation}
\rho=\rho_0 x^3,\qquad
p=K\,\phi(x),\qquad
\phi(x)\equiv x(2x^2-3)\sqrt{1+x^2}+3\sinh^{-1}(x)
\end{equation}
together with the length and mass scales
\begin{equation}
a \equiv \sqrt{\frac{K}{4\pi G\,\rho_0^2}},\qquad
m_0 \equiv 4\pi \rho_0 a^3,
\label{eq:scales}
\end{equation}
and introducing the variables
\begin{equation}
\bar r \equiv \frac{r}{a},\qquad
\bar m \equiv \frac{m}{m_0}.
\end{equation}
Then, the structure equations become
\begin{align}
\frac{\dd x}{\dd \bar r} &= -\frac{\bar m\, x^3}{\bar r^2}\left(\frac{\dd \phi}{\dd x}\right)^{-1},
\label{eq:dxdr_dimless}\\
\frac{\dd \bar m}{\dd \bar r} &= \bar r^2 x^3,
\label{eq:dmdr_dimless}
\end{align}
with $x(0)=x_c=(\rho_c/\rho_0)^{1/3}$ and $\bar m(0)=0$.
The solutions of this set of equations  define the reference sequence $(M_0(\rho_c),R_0(\rho_c))$ used below \citep{Chandrasekhar1939,ShapiroTeukolsky1983}.

\section{Rotation-induced anisotropy in Newtonian white dwarfs}
\label{sec:rotation_anisotropy}

We incorporate uniform rotation through a one-dimensional model mapping that replaces the angle-averaged
centrifugal support by an effective pressure anisotropy \citep{Tassoul1978,Chandrasekhar1969}. In this construction, the gravitational sector and the mass
continuity remain spherically symmetric; rotation enters only through the mapped term derived below.
\subsection{Angle-averaged radial balance}
\label{subsec:rot_euler}

In a uniformly rotating frame, the Newtonian Euler equation reads
\begin{equation}
\frac{1}{\rho}\nabla p = -\nabla \Phi + \Omega^2 \varpi\,\hat{\bm{\varpi}},
\end{equation}
where $\varpi=r\sin\theta$ \citep{Tassoul1978}. Projecting onto the spherical radial direction gives
\begin{equation}
\frac{1}{\rho}\frac{\partial p}{\partial r}
=
-\frac{\partial \Phi}{\partial r}
+
\Omega^2 r\sin^2\theta.
\end{equation}
We then replace the latitude dependence of the centrifugal term by its angular average
$\langle \sin^2\theta \rangle=2/3$, obtaining the effective one-dimensional balance
\begin{equation}
\frac{\mathrm{d}p}{\mathrm{d}r}
=
-\rho\,\frac{\mathrm{d}\Phi}{\mathrm{d}r}
+
\frac{2}{3}\rho\,\Omega^2 r.
\label{eq:avg_balance}
\end{equation}
\citep{Tassoul1978,Chandrasekhar1969}.
\subsection{Mapping to anisotropic hydrostatic equilibrium}
\label{subsec:rot_mapping}

Introduce radial and tangential pressures $(p_r,p_t)$ and the anisotropy function
\begin{equation}
\Delta(r)\equiv p_t(r)-p_r(r).
\end{equation}
The Newtonian anisotropic hydrostatic equation is
\begin{equation}
\frac{\mathrm{d}p_r}{\mathrm{d}r}
=
-\rho\,\frac{\mathrm{d}\Phi}{\mathrm{d}r}
+
\frac{2}{r}\Delta(r).
\label{eq:aniso_hse}
\end{equation}
\citep{BowersLiang1974}.
Comparing Eqs.~\eqref{eq:avg_balance} and \eqref{eq:aniso_hse}, we identify the rotation-induced effective anisotropy
\begin{equation}
\Delta_{\rm rot}(r)=\frac{1}{3}\rho(r)\Omega^2 r^2,
\label{eq:delta_rot}
\end{equation}
which is regular at the origin \citep{BowersLiang1974}. In this reduced model, the barotropic equation of state is applied to the radial pressure,
$p_r=p(\rho)$ (Sec.~\ref{subsec:baseline_eos}), while the tangential component follows as $p_t=p_r+\Delta_{\rm rot}$.
\subsection{Structure equations with rotation-induced anisotropy}
\label{subsec:rot_structure}

The mass equation remains unchanged,
\begin{equation}
\frac{\mathrm{d}m}{\mathrm{d}r} = 4\pi r^2\rho(r).
\label{eq:mass_cont_rot}
\end{equation}
To close the one-dimensional system, we adopt the spherical potential gradient
$\mathrm{d}\Phi/\mathrm{d}r = Gm(r)/r^2$, consistent with the spherically symmetric gravitational sector retained in the
reduced model. The radial force balance becomes
\begin{equation}
\frac{\mathrm{d}p_r}{\mathrm{d}r}
=
-\frac{Gm(r)\rho(r)}{r^2}
+
\frac{2}{3}\rho(r)\Omega^2 r,
\label{eq:hse_rot_final}
\end{equation}
with $p_r=p(\rho)$ from Sec.~\ref{subsec:baseline_eos}.
\subsection{Rotation diagnostics and admissibility}
\label{subsec:rot_applicability}

For each configuration we monitor the dimensionless rotation parameter
\begin{equation}
\epsilon \equiv \frac{\Omega^2 R^3}{GM},
\label{eq:epsilon_rot}
\end{equation}
and enforce the mass-shedding (Kepler) bound evaluated on the rotating $(M,R)$,
\begin{equation}
\Omega < \Omega_K(M,R)\equiv \sqrt{\frac{GM}{R^3}}
\qquad \Longleftrightarrow \qquad
\epsilon < 1.
\label{eq:kepler_rot}
\end{equation}
\citep{Tassoul1978,Chandrasekhar1969}.
We also track the interior ratio $\Delta_{\rm rot}/p_r$ as a practical smallness indicator for the mapped term.

\subsection{Numerical implementation and robustness checks}
\label{subsec:numerical_details}

We integrate Eqs.~\eqref{eq:mass_cont_rot} and \eqref{eq:hse_rot_final} outward from the regular centre,
\[
m(0)=0,\qquad \rho(0)=\rho_c .
\]
The calculation is performed in the dimensionless variables introduced in
Sec.~\ref{subsec:baseline_dimless},
\[
\rho=\rho_0 x^3,\qquad p_r=K\phi(x),\qquad r=a\bar r,\qquad m=m_0\bar m .
\]
In these variables, the mass equation becomes
\begin{equation}
\frac{d\bar m}{d\bar r}=\bar r^2 x^3 ,
\end{equation}
while the radial equilibrium equation with the mapped centrifugal term is
\begin{equation}
\frac{dx}{d\bar r}
=
x^3
\left[
-\frac{\bar m}{\bar r^2}
+
\beta_\Omega \bar r
\right]
\left(\frac{d\phi}{dx}\right)^{-1},
\qquad
\beta_\Omega\equiv \frac{\Omega^2}{6\pi G\rho_0}.
\end{equation}
The first term in brackets is the usual Newtonian gravitational contribution, and the second
term is the dimensionless form of the angle-averaged centrifugal support in
Eq.~\eqref{eq:hse_rot_final}. The non-rotating Chandrasekhar sequence is recovered by
setting $\Omega=0$.

The integration is started at a small radius,
\[
\bar r_0=10^{-8},
\]
using the regular central expansion. To leading order,
\begin{equation}
\bar m(\bar r_0)=\frac{x_c^3}{3}\bar r_0^3 ,
\end{equation}
and
\begin{equation}
x(\bar r_0)
=
x_c+
\frac{1}{2}
\frac{x_c^3}{\phi'(x_c)}
\left(
\beta_\Omega-\frac{x_c^3}{3}
\right)
\bar r_0^2 ,
\end{equation}
where $x_c=(\rho_c/\rho_0)^{1/3}$. This initialization avoids the coordinate singularity at
the origin and gives a smooth starting point for both rotating and non-rotating models.

The surface is defined numerically by the first point at which the Chandrasekhar parameter
falls to
\begin{equation}
x(\bar r)=x_{\rm stop},\qquad x_{\rm stop}=10^{-10}.
\end{equation}
This corresponds to $p(R)\approx 0$ and $\rho(R)\approx 0$ within the adopted numerical
tolerance. The dimensional mass and radius are then obtained from
\[
M=m_0\bar m(\bar R),\qquad R=a\bar R .
\]

For global scans we sample
\[
\log_{10}\rho_c\in[6,11]
\]
in steps of $0.25$ dex. For each central density, the rotation rate is assigned through the
proxy $f=\Omega/\Omega_{K,0}(\rho_c)$, where $\Omega_{K,0}$ is computed from the
corresponding non-rotating reference model. Since this proxy is tied to the baseline
configuration, the final admissibility diagnostics are evaluated on the rotating solution
itself. For each model we store $(M,R)$, the interior profiles, and the diagnostics
$\epsilon$ and $\Omega/\Omega_K(M,R)$.

As robustness checks, we repeated representative integrations with tighter solver tolerances,
larger maximum integration radii, and locally refined central-density sampling near the
high-density end. These tests did not change the tabulated readout or the reported trends
within the numerical precision relevant for the figures and tables below.

\paragraph{Remark on the surface behavior.}
Equation~\eqref{eq:delta_rot} does not enforce $\Delta_{\rm rot}(R)=0$. When desired, we apply a smooth taper
\begin{equation}
\Delta_{\rm rot}(r)\rightarrow \Delta_{\rm rot}(r)\,T(r),\qquad
T(r)=1-\left(\frac{r}{R}\right)^n,
\end{equation}
with $n=2$ unless stated otherwise. We use this only as a sensitivity check. The resulting
changes in $(M,R)$ remain at the level of the numerical resolution for the sequences
reported below, so the untapered form is used as the default reduced model.

\section{Applicability diagnostics and reference formulations}
\label{sec:applicability_methods}

This section defines the diagnostic quantities used to interpret the one-dimensional
reduced model and records the reference formulations used for comparison. The aim is not
to establish a complete stability theory for rotating white dwarfs, but to specify the
operational readout and the range in which the angle-averaged mapping can be used as a
controlled approximation.

\subsection{Operational meaning of the high-density readout}
\label{subsec:stability_scope}
\label{subsec:limiting_readout}

For the cold Newtonian Chandrasekhar equation of state used here, the non-rotating mass
sequence approaches an asymptotic value at large central density
\citep{Chandrasekhar1931,Chandrasekhar1939,ShapiroTeukolsky1983}. In a finite numerical
scan this appears as a high-density saturation region rather than as a sharply resolved
interior maximum of $M(\rho_c)$. We use the same interpretation for the rotating reduced
sequences. Thus, when the largest value of $M(\rho_c;f)$ occurs at the upper edge of the
sampled density interval, it is not taken as a stability boundary, but only as a saturation
readout within the explored range.

For each fixed-$f$ family, we define
\begin{equation}
M_{\rm lim}(f)\equiv \max_{\rho_c} M(\rho_c;f),\qquad
\rho_{c,\rm lim}(f)\equiv \arg\max_{\rho_c} M(\rho_c;f),\qquad
R_{\rm lim}(f)\equiv R\!\left(\rho_{c,\rm lim}(f);f\right).
\label{eq:limiting_def}
\end{equation}
Operationally, $M_{\rm lim}$ is the discrete maximum of
$M(\log_{10}\rho_c;f)$ on the scanned grid. In the present calculations this maximum
occurs at $\log_{10}\rho_c=11$ for all reported fixed-$f$ sequences. The quantities
$(M_{\rm lim},R_{\rm lim})$ should therefore be read as high-density saturation estimates
for the adopted Newtonian microphysics and density interval, not as exact maximum-mass
configurations of rotating white dwarfs.

This distinction fixes the scope of the analysis. We do not perform a radial perturbation
calculation and do not derive a full stability criterion for rotating equilibria. The limiting
readout is used only to compare the fixed-$f$ sequences on a common basis. The range of
use of the one-dimensional rotation mapping is assessed separately through the diagnostics
introduced below.

\subsection{Physical constraints and empirical context}
\label{subsec:newtonian_rotation_constraints}

Uniform rotation is constrained first by mass shedding. In the present reduced model this
condition must be evaluated on the rotating configuration, not on the non-rotating reference
model. We therefore use $\Omega/\Omega_K(M,R)$, or equivalently $\epsilon$, as the
sub-Keplerian consistency check \citep{Tassoul1978}.

At higher rotation rates, non-axisymmetric instabilities may also become relevant. Classical
analyses often characterize their onset through $T/|W|$, with benchmark values of order
$0.1$--$0.3$ for secular or dynamical bar-mode instabilities
\citep{Chandrasekhar1969,LaiRasioShapiro1993,OstrikerPeebles1973}. We use these values
only as qualitative context and do not impose a $T/|W|$ cut, since the present
one-dimensional model does not resolve the stellar shape.

The observed spin distribution of WDs further supports this separation between
slow-rotation survey models and fast-rotation configurations requiring multidimensional
treatment \citep{Hermes2017ApJS,Warner1995CVs,Tassoul1978,Hachisu1986ApJS}.

\subsection{Applicability diagnostics}
\label{subsec:applicability_diagnostics}

The scan parameter $f\equiv \Omega/\Omega_{K,0}(\rho_c)$ is useful for constructing
sequences, but it is defined with respect to the non-rotating baseline. Physical
admissibility must therefore be checked on the rotating solution itself. For every computed
model we evaluate
\begin{equation}
\Omega_K(M,R)=\sqrt{\frac{GM}{R^3}},\qquad
\epsilon \equiv \frac{\Omega^2 R^3}{GM},
\label{eq:OmegaK_eps}
\end{equation}
and record $\max(\Omega/\Omega_K)$ and $\max(\epsilon)$ along each fixed-$f$ family
(Table~\ref{tab:summary}). The condition $\Omega/\Omega_K<1$, equivalently
$\epsilon<1$, is used as the mass-shedding consistency requirement \citep{Tassoul1978}.

A second diagnostic measures the relative size of the mapped anisotropic term in the bulk
interior,
\begin{equation}
\mathcal{A}_{\eta}\equiv \max_{p_r/p_c\ge \eta}\left(\frac{\Delta_{\rm rot}}{p_r}\right),
\qquad \eta=10^{-2}.
\label{eq:Aeta_def_env_app}
\end{equation}
The cutoff excludes the low-pressure surface layer, where $p_r\to0$ and
$\Delta_{\rm rot}/p_r$ is dominated by a boundary effect rather than by the bulk force
balance. The resulting $\mathcal{A}_{10^{-2}}$ values are reported in
Table~\ref{tab:summary}. Varying the cutoff within
$\eta\in[10^{-3},10^{-2}]$ does not change the qualitative conclusions:
$\mathcal{A}_\eta$ remains below unity and increases monotonically with $f$ over the
scanned range.

\subsection{Numerical extraction and readout}
\label{subsec:stability_numerics}

Equilibrium sequences are computed on the logarithmic grid
$\rho_c\in[10^{6},10^{11}]\,{\rm g\,cm^{-3}}$ with a step of $0.25$ dex. For each fixed-$f$
family, we extract $(M_{\rm lim},R_{\rm lim},\rho_{c,\rm lim})$ using
Eq.~\eqref{eq:limiting_def}, and record the corresponding values of
$\max(\Omega/\Omega_K)$, $\max(\epsilon)$, and $\mathcal{A}_{10^{-2}}$
(Table~\ref{tab:summary}).

In the explored range $f\le0.35$, the limiting mass and radius increase monotonically with
$f$, while all reported families remain sub-Keplerian. The increase in the limiting mass is
modest, reaching the percent level for the largest retained rotation proxy. The
bulk-interior diagnostic $\mathcal{A}_{10^{-2}}$ also remains below unity, indicating that
the mapped anisotropic term stays subdominant to the radial pressure away from the surface
layer.

As numerical checks, we repeated representative integrations with tighter solver tolerances
and locally refined the density sampling near the high-density end. These tests did not
change the tabulated readout or the trends with the rotation proxy within the numerical
precision relevant for the results below.

\subsection{Reference axisymmetric formulation}
\label{subsec:benchmark_2D}

The one-dimensional rotation--anisotropy mapping is useful for rapid surveys, but it does
not include the axisymmetric deformation of rotating configurations. For reference, we
recall the corresponding Newtonian formulation for a uniformly rotating barotrope with the
same microphysics \citep{Tassoul1978,Hachisu1986ApJS,OstrikerBodenheimer1968}. This
formulation is not solved here; it only identifies the natural self-consistent benchmark for
future calibration of the reduced model.

In Newtonian gravity, an axisymmetric, stationary, uniformly rotating barotrope satisfies
the first integral
\begin{equation}
H(\rho) + \Phi - \frac{1}{2}\Omega^2 \varpi^2 = C,
\label{eq:2D_first_integral}
\end{equation}
where $\varpi$ is the cylindrical radius, $\Phi$ is the gravitational potential, and
\begin{equation}
H(\rho)\equiv \int_{0}^{\rho}\frac{dp}{\rho'}
\label{eq:enthalpy_def}
\end{equation}
is the specific enthalpy for the barotropic Chandrasekhar equation of state,
$p=p(\rho)$ \citep{Tassoul1978,Hachisu1986ApJS}. The potential is determined by
Poisson's equation,
\begin{equation}
\nabla^2\Phi = 4\pi G\rho.
\label{eq:poisson}
\end{equation}
Equations~\eqref{eq:2D_first_integral}--\eqref{eq:poisson} define the self-consistent
axisymmetric equilibrium once $\Omega$ and $\rho_c$ are specified, with the surface fixed
by a condition such as $H=0$ \citep{Tassoul1978}.

A standard numerical route is a self-consistent field iteration on a cylindrical grid
$(\varpi,z)$, in which the gravitational potential, enthalpy field, and density distribution
are updated until convergence \citep{OstrikerBodenheimer1968,Hachisu1986ApJS}. Such a
calculation yields the equatorial and polar radii, $R_{\rm eq}$ and $R_{\rm p}$, and therefore
the stellar deformation. For comparison with a one-dimensional model, it is useful to define
the volume-equivalent radius
\begin{equation}
R_{\rm vol}\equiv \left(\frac{3V}{4\pi}\right)^{1/3},
\label{eq:Rvol_def}
\end{equation}
where $V$ is the stellar volume. The oblateness may be measured by
$\mathcal{O}\equiv R_{\rm eq}/R_{\rm p}$, while mass-shedding proximity can be estimated
with $\Omega_K\simeq \sqrt{GM/R_{\rm eq}^3}$ \citep{Chandrasekhar1969,Tassoul1978}.

For a direct calibration, the reduced-model outputs $(M_{\rm 1D},R_{\rm 1D})$ could be
compared with the axisymmetric outputs $(M_{\rm 2D},R_{\rm vol,2D})$ through
\begin{equation}
\delta_M \equiv \frac{M_{\rm 1D}-M_{\rm 2D}}{M_{\rm 2D}},\qquad
\delta_R \equiv \frac{R_{\rm 1D}-R_{\rm vol,2D}}{R_{\rm vol,2D}}.
\label{eq:deltaMR_def}
\end{equation}
These quantities would measure the error introduced by the angle-averaged reduction as the
rotation rate increases \citep{Hachisu1986ApJS}. In the present work, the
quasi-two-dimensional reconstruction in Sec.~\ref{subsec:quasi2D_check} is used only as an
auxiliary consistency check, not as a replacement for this self-consistent benchmark.

\section{Newtonian reduced-model results}
\label{sec:results}

This section presents the numerical results of the reduced model. We first show representative
interior profiles, then discuss the global sequences and high-density readout. We then apply
the multi-criterion applicability envelope and give the quasi-two-dimensional consistency check.

\subsection{Interior profiles}
\label{subsec:results_interior_profiles}

\begin{figure}
\centering
\includegraphics[width=0.6\textwidth]{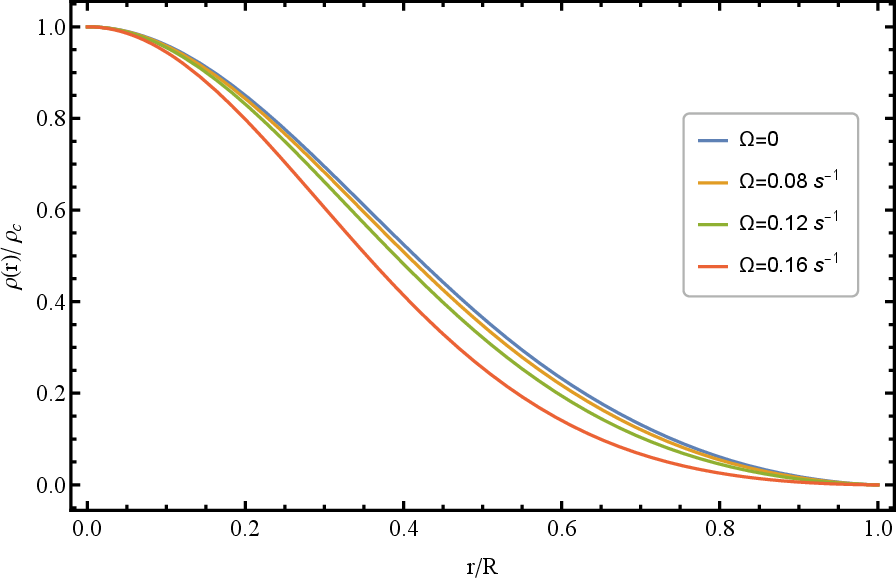}
\caption{Normalized density profile $\rho(r)/\rho_c$ as a function of $r/R$ for several uniform rotation rates $\Omega$
(see legend). Solutions are obtained from Eqs.~\eqref{eq:mass_cont_rot}--\eqref{eq:hse_rot_final}, with rotation
incorporated through Eq.~\eqref{eq:delta_rot}.}
\label{fig:rho_rho}
\end{figure}

\begin{figure}
\centering
\includegraphics[width=0.6\textwidth]{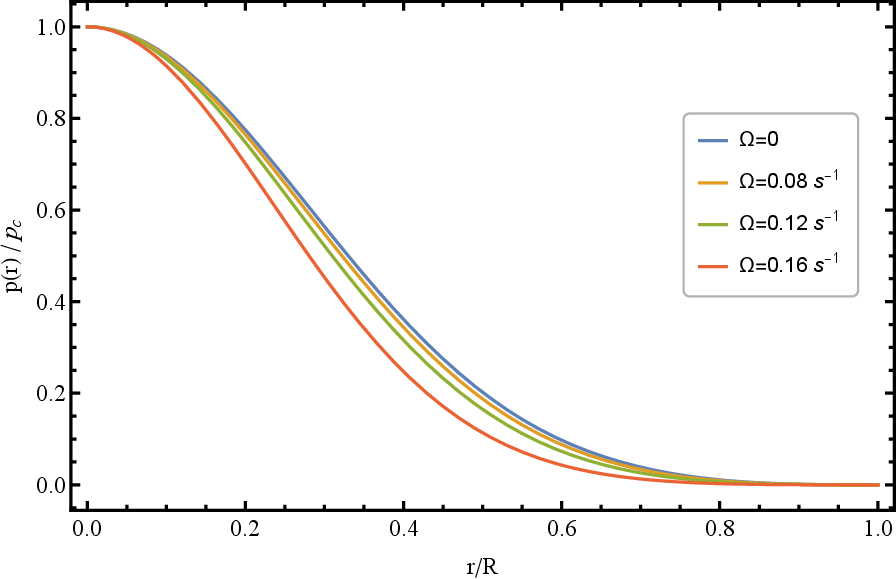}
\caption{Normalized radial pressure profile $p_r(r)/p_c$ versus $r/R$ for the same set of rotation rates as in
Fig.~\ref{fig:rho_rho}, where $p_c\equiv p_r(0)$.}
\label{fig:p_p}
\end{figure}

\begin{figure}
\centering
\includegraphics[width=0.6\textwidth]{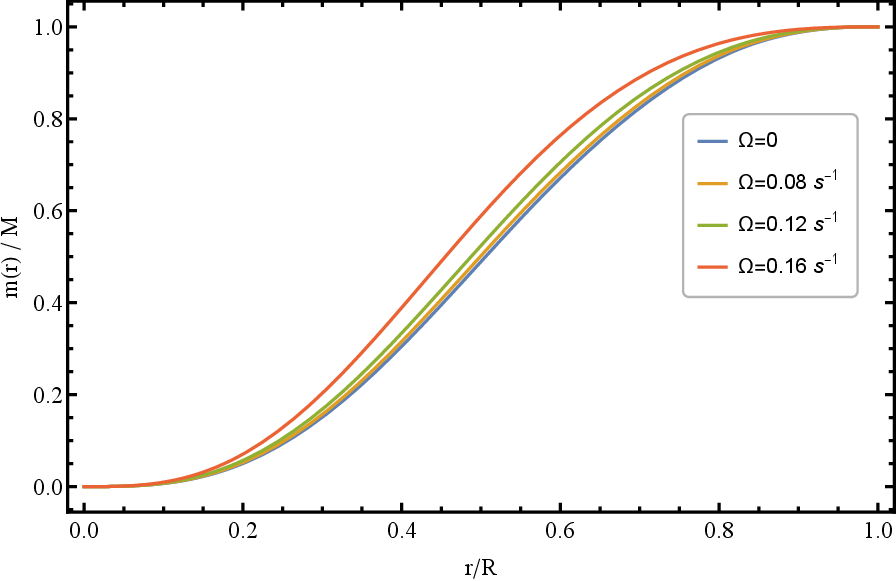}
\caption{Cumulative enclosed mass fraction $m(r)/M$ as a function of $r/R$ for different rotation rates.}
\label{fig:m_m}
\end{figure}

\begin{figure}[H]
\centering
\includegraphics[width=0.6\textwidth]{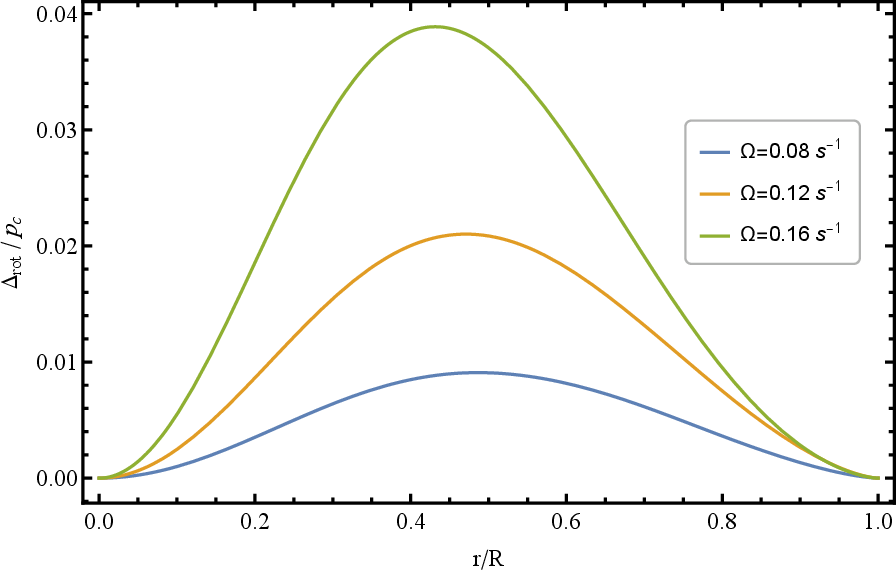}
\caption{Rotation-induced effective anisotropy $\Delta_{\rm rot}(r)/p_c$ versus $r/R$ for several rotation rates.}
\label{fig:delta}
\end{figure}

Figures~\ref{fig:rho_rho}--\ref{fig:delta} show representative dimensionless profiles as
functions of $r/R$ for several uniform rotation rates. As the rotation rate increases, the
density and radial-pressure profiles become slightly broader, while the enclosed-mass
fraction changes consistently with the additional angle-averaged centrifugal support. The
effective anisotropy profile remains regular at the centre, as required by
Eq.~\eqref{eq:delta_rot}, and reaches its largest values inside the star rather than at the
origin.

\FloatBarrier

\subsection{Global sequences and high-density readout}
\label{subsec:results_global_sequences}

We compute the global sequences using the rotation proxy
$f=\Omega/\Omega_{K,0}(\rho_c)$ introduced above, with
$\rho_c\in[10^{6},10^{11}]\,{\rm g\,cm^{-3}}$ sampled in steps of $0.25$ dex and
$f=\{0,0.15,0.25,0.35\}$. Since this proxy is defined with respect to the non-rotating
reference model, the sub-Keplerian diagnostics are always evaluated on the corresponding
rotating configurations.

\begin{figure}
\centering
\includegraphics[width=0.6\textwidth]{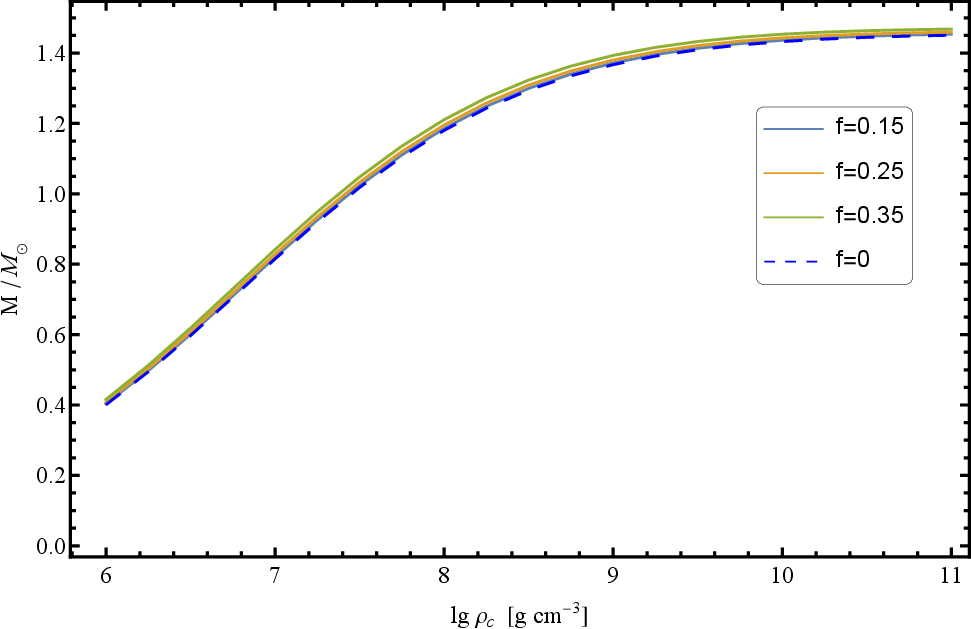}
\caption{Total mass $M$ (in units of $M_\odot$) as a function of the central density $\rho_c$ for rotation rates
parameterized by $f\equiv \Omega/\Omega_{K,0}(\rho_c)$ (legend). The dashed curve ($f=0$) shows the baseline sequence.}
\label{fig:sprint_fig1}
\end{figure}

\begin{figure}
\centering
\includegraphics[width=0.6\textwidth]{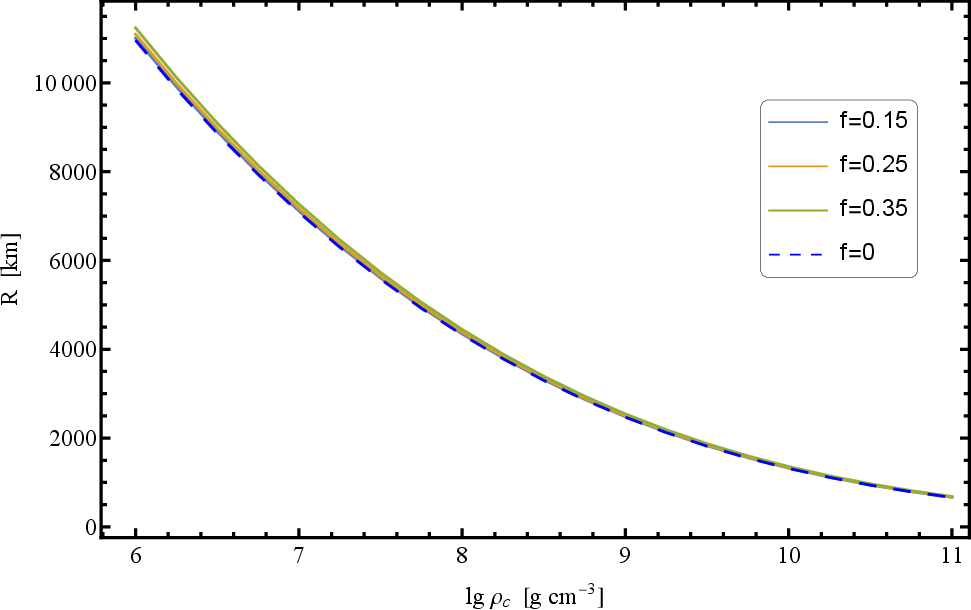}
\caption{Radius $R$ (km) as a function of $\rho_c$ for the same set of rotation fractions as in
Fig.~\ref{fig:sprint_fig1}.}
\label{fig:sprint_fig2}
\end{figure}

\begin{figure}
\centering
\includegraphics[width=0.6\textwidth]{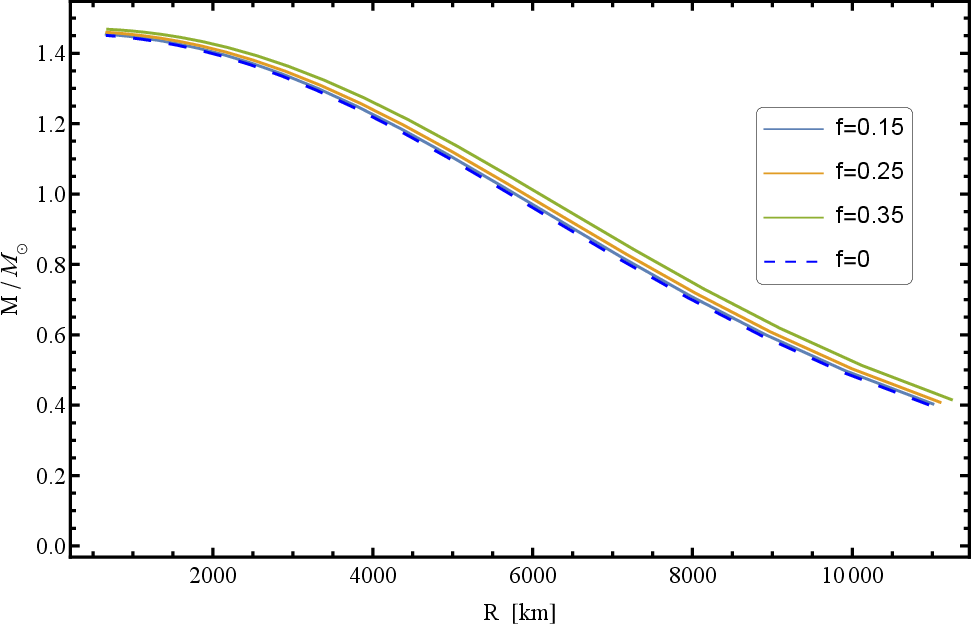}
\caption{Mass--radius relation $M(R)$ for sequences at fixed $f$, compared to the non-rotating baseline ($f=0$).}
\label{fig:sprint_fig3}
\end{figure}

Figures~\ref{fig:sprint_fig1}--\ref{fig:sprint_fig3} show a systematic upward shift of the
mass and radius as the rotation proxy increases. The effect remains modest over the
retained range of $f$, but it is monotonic. The high-density readout is interpreted in the
operational sense defined in Sec.~\ref{subsec:limiting_readout}; in the present grid, the
discrete maximum occurs at $\log_{10}\rho_c=11$ for all fixed-$f$ sequences.

\begin{table}[t]
\centering
\caption{High-density saturation readout for the fixed-$f$ sequences shown in
Figs.~\ref{fig:sprint_fig1}--\ref{fig:sprint_fig3}. The quantities are evaluated at
$\log_{10}\rho_c=11$. The mass increase is given relative to the non-rotating baseline, and the last
three columns list the sub-Keplerian and bulk-interior applicability diagnostics.}
\label{tab:summary}
\begin{tabular}{c c c c c c c c}
\hline\hline
$f$ & $M_{\rm lim}\,[M_\odot]$ & $R_{\rm lim}\,[{\rm km}]$ & $\log_{10}\rho_{c,\rm lim}$ &
$\Delta M_{\rm lim}[\%]$ & $\max(\Omega/\Omega_K)$ & $\max(\epsilon)$ & $\mathcal{A}_{10^{-2}}$ \\
\hline
0.00 & 1.45072 & 662.771 & 11.00 & 0.0000 & ---     & ---      & 0 \\
0.15 & 1.45384 & 666.374 & 11.00 & 0.2153 & 0.15106 & 0.0228198 & 0.05147 \\
0.25 & 1.45947 & 673.117 & 11.00 & 0.6036 & 0.25511 & 0.0650801 & 0.14432 \\
0.35 & 1.46813 & 684.176 & 11.00 & 1.2002 & 0.36491 & 0.133158  & 0.28436 \\
\hline\hline
\end{tabular}
\end{table}

Table~\ref{tab:summary} shows that the largest retained proxy, $f=0.35$, produces a
$1.2\%$ increase in the high-density mass readout and a corresponding increase in the
radius. All reported sequences remain sub-Keplerian, with
$\max(\Omega/\Omega_K)<1$ and $\max(\epsilon)<1$. The bulk-interior measure
$\mathcal{A}_{10^{-2}}$ also remains below unity, although it increases with $f$, indicating
that the mapped anisotropic term is still subdominant in the retained range.

\FloatBarrier

\subsection{Multi-criterion applicability envelope for the rotation proxy}
\label{subsec:applicability_envelope}

For parameter sweeps we use the proxy
\begin{equation}
f \equiv \frac{\Omega}{\Omega_{K,0}(\rho_c)},\qquad 
\Omega_{K,0}(\rho_c)\equiv \sqrt{\frac{G M_0(\rho_c)}{R_0(\rho_c)^3}},
\label{eq:f_def}
\end{equation}
defined with respect to the \emph{non-rotating} baseline model at the same $\rho_c$. Because this reference differs
from the rotating configuration, $f$ is not bounded \emph{a priori}. The physically admissible and model-valid range
must therefore be established by explicit constraints evaluated on the rotating solutions.

\paragraph{Hard physical bound (mass shedding).}
For each fixed-$f$ family we compute the Kepler frequency from the rotating $(M,R)$,
$\Omega_K(M,R)=\sqrt{GM/R^3}$, and define the family-level diagnostic
\begin{equation}
K_1(f)\equiv \max_{\rho_c}\left(\frac{\Omega}{\Omega_K(M,R)}\right),
\label{eq:K1_def}
\end{equation}
where the maximum is taken along the scanned $\rho_c$ grid. The condition $K_1(f)<1$ enforces sub-Keplerian rotation
(equivalently $\max(\epsilon)<1$ with $\epsilon=\Omega^2R^3/(GM)$) \citep{Tassoul1978}.

\paragraph{Bulk force-balance smallness.}
To test whether the mapped centrifugal contribution remains perturbative in the radial balance, we compare the rotation
term $(2/3)\rho\,\Omega^2 r$ to the Newtonian gravitational term $Gm\rho/r^2$ and introduce
\begin{equation}
\chi(r)\equiv \frac{\frac{2}{3}\rho\,\Omega^2 r}{Gm\rho/r^2}
=\frac{2}{3}\frac{\Omega^2 r^3}{Gm(r)}.
\label{eq:chi_r_def}
\end{equation}
We summarize the bulk interior through
\begin{equation}
\chi_{10^{-2}} \equiv \max_{p_r/p_c\ge 10^{-2}}\chi(r),
\label{eq:chi_eta_def}
\end{equation}
and define
\begin{equation}
K_2(f)\equiv \frac{\chi_{10^{-2}}}{\chi_\star}, \qquad \chi_\star=0.2.
\label{eq:K2_def}
\end{equation}
The value $\chi_\star=0.2$ is used as a conservative working threshold for keeping the
rotational term at the $\lesssim20\%$ level of the gravitational term in the bulk interior.

\paragraph{Bulk effective-anisotropy smallness.}
A complementary diagnostic monitors the mapped anisotropic term relative to the radial pressure,
\begin{equation}
\mathcal{A}_{10^{-2}}\equiv \max_{p_r/p_c\ge 10^{-2}}\left(\frac{\Delta_{\rm rot}}{p_r}\right),
\qquad \Delta_{\rm rot}(r)=\frac{1}{3}\rho(r)\Omega^2r^2,
\label{eq:Aeta_def_env}
\end{equation}
and we introduce
\begin{equation}
K_3(f)\equiv \frac{\mathcal{A}_{10^{-2}}}{A_\star},\qquad A_\star=0.3.
\label{eq:K3_def}
\end{equation}
The value $A_\star=0.3$ is again a conservative working threshold. It is not a universal
stability limit; it only enforces that the mapped anisotropic term remains subdominant to
the bulk radial pressure.

\begin{figure}
\centering
\includegraphics[width=0.6\textwidth]{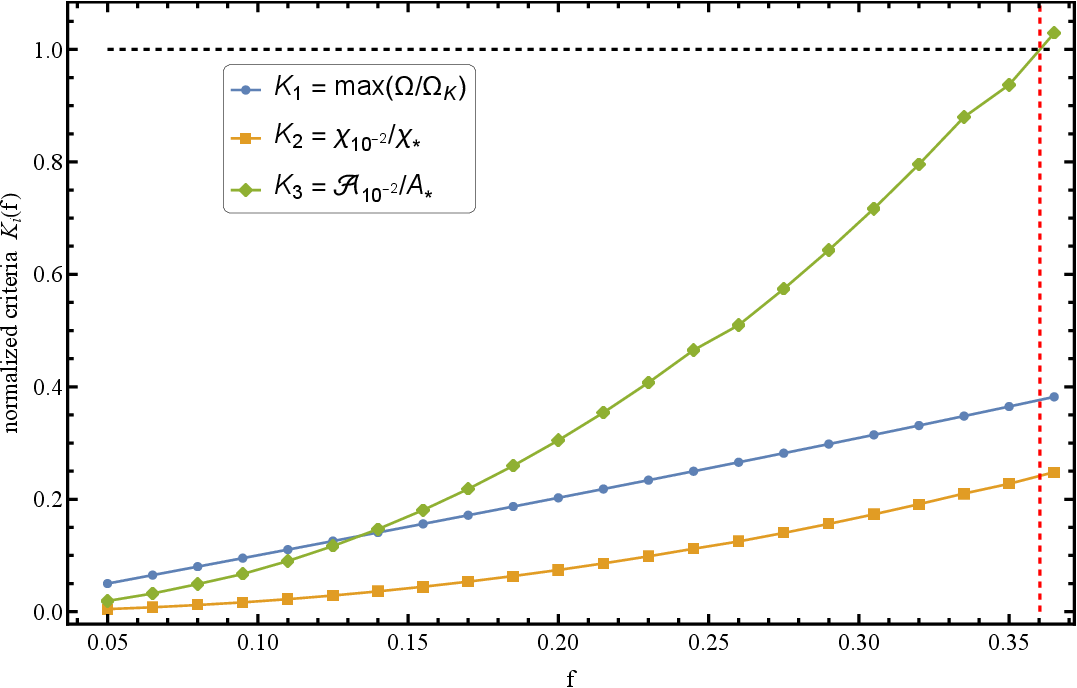}
\caption{Normalized applicability criteria for the rotation proxy $f$.
We show $K_1=\max(\Omega/\Omega_K)$ (sub-Keplerian proximity), $K_2=\chi_{10^{-2}}/\chi_\star$ with
$\chi(r)=(2/3)\Omega^2 r^3/[Gm(r)]$ and $\chi_\star=0.2$ (bulk force-balance smallness), and
$K_3=\mathcal{A}_{10^{-2}}/A_\star$ with $\mathcal{A}_{10^{-2}}=\max_{p_r/p_c\ge10^{-2}}(\Delta_{\rm rot}/p_r)$ and
$A_\star=0.3$ (bulk anisotropy smallness). The horizontal dashed line marks the common threshold $K_i=1$. The vertical
dashed line indicates the recommended slow-rotation bound $f_{\rm rec}$ defined by the earliest threshold crossing
among the three criteria.}
\label{fig:criteria_envelope}
\end{figure}

\paragraph{Recommended slow-rotation domain.}
The three criteria are placed on a common scale by requiring
\begin{equation}
K_i(f) < 1\qquad (i=1,2,3).
\label{eq:Ki_threshold}
\end{equation}
The recommended slow-rotation bound is defined by the intersection of the three admissible
ranges,
\begin{equation}
f \le f_{\rm rec}\equiv \min\!\left(f_{\max}^{(1)},\,f_{\max}^{(2)},\,f_{\max}^{(3)}\right),
\qquad K_i\!\left(f_{\max}^{(i)}\right)=1,
\label{eq:frec_def}
\end{equation}
where each $f_{\max}^{(i)}$ is obtained by interpolation between adjacent sampled values of
$f$ with $\Delta f=0.015$. Figure~\ref{fig:criteria_envelope} shows the resulting envelope.
For the present scan, the effective-anisotropy smallness criterion is the most restrictive,
while the sub-Keplerian condition remains comfortably satisfied.

\FloatBarrier

\subsection{Quasi-two-dimensional consistency check}
\label{subsec:quasi2D_check}

As an auxiliary check, we performed a quasi-two-dimensional reconstruction for a small
representative set of models. This is not a self-consistent axisymmetric equilibrium
calculation. It uses the spherical background potential of the corresponding non-rotating
model and adds the centrifugal contribution at the level of the enthalpy distribution. The
purpose is only to test whether the one-dimensional readout remains close to a simple
axisymmetric reconstruction in the slow-rotation regime.

For each selected central density, we computed the corresponding non-rotating spherical
background and used its gravitational potential as a reference field. We then constructed an
approximate axisymmetric enthalpy distribution in cylindrical coordinates by adding the
centrifugal term. From the resulting density field we evaluated an approximate
two-dimensional mass, $M_{2D}^{\mathrm{approx}}$, and a volume-equivalent radius,
$R_{\mathrm{vol},2D}^{\mathrm{approx}}$. These quantities were compared with the
one-dimensional outputs $(M_{1D},R_{1D})$ through
\begin{equation}
\delta M \equiv \frac{M_{1D}-M_{2D}^{\mathrm{approx}}}{M_{2D}^{\mathrm{approx}}},
\qquad
\delta R \equiv \frac{R_{1D}-R_{\mathrm{vol},2D}^{\mathrm{approx}}}{R_{\mathrm{vol},2D}^{\mathrm{approx}}}.
\label{eq:deltaMR_quasi2D}
\end{equation}

The comparison was performed for
\[
\rho_c=10^{6},\,10^{8},\,10^{10},\,10^{11}\ {\rm g\,cm^{-3}},
\qquad
q\equiv \Omega/\Omega_K=\{0.1,0.2,0.3\}.
\]
Here $q$ denotes the rotation fraction used only in the quasi-two-dimensional reconstruction
and should not be confused with the sequence parameter $f$ used above. Over this set, the
mass discrepancy remains at the $10^{-3}$ level, while the radius discrepancy remains at
the percent level. The maximum values are
\[
|\delta M|\lesssim 3.03\times 10^{-3},\qquad
|\delta R|\lesssim 1.12\times 10^{-2}.
\]
The discrepancies grow mainly with the rotation fraction, as expected when the neglected
axisymmetric deformation becomes more important.

\begin{figure}
\centering
\includegraphics[width=0.75\textwidth]{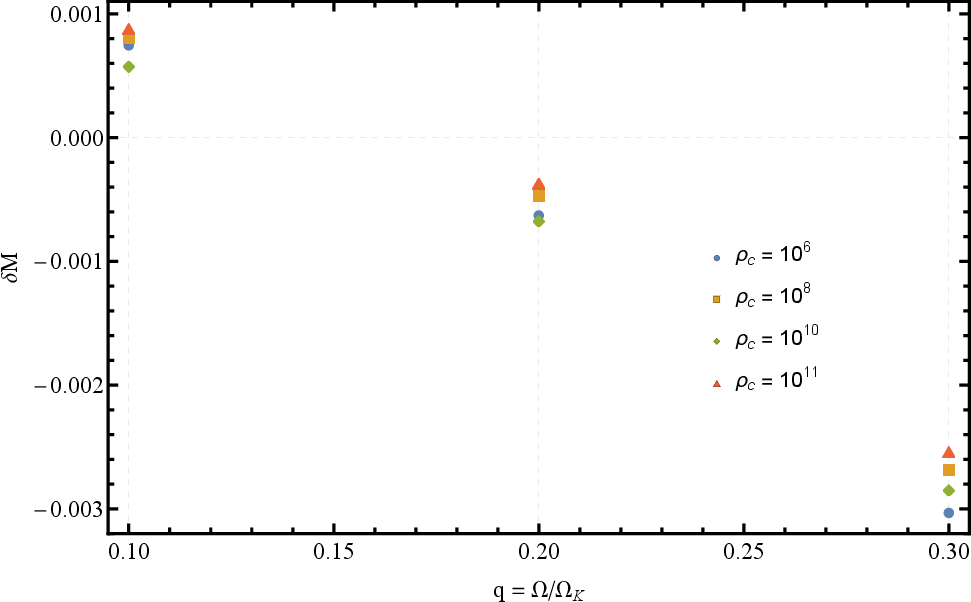}
\caption{Relative mass discrepancy between the one-dimensional reduced model and the quasi-two-dimensional reconstruction as a function of the rotation fraction $q=\Omega/\Omega_K$, for the sampled set of central densities $\rho_c=10^6,10^8,10^{10},10^{11}\,{\rm g\,cm^{-3}}$. The discrepancy remains at the $10^{-3}$ level over the tested range and increases systematically with $q$.}
\label{fig:dM}
\end{figure}

\begin{figure}
\centering
\includegraphics[width=0.75\textwidth]{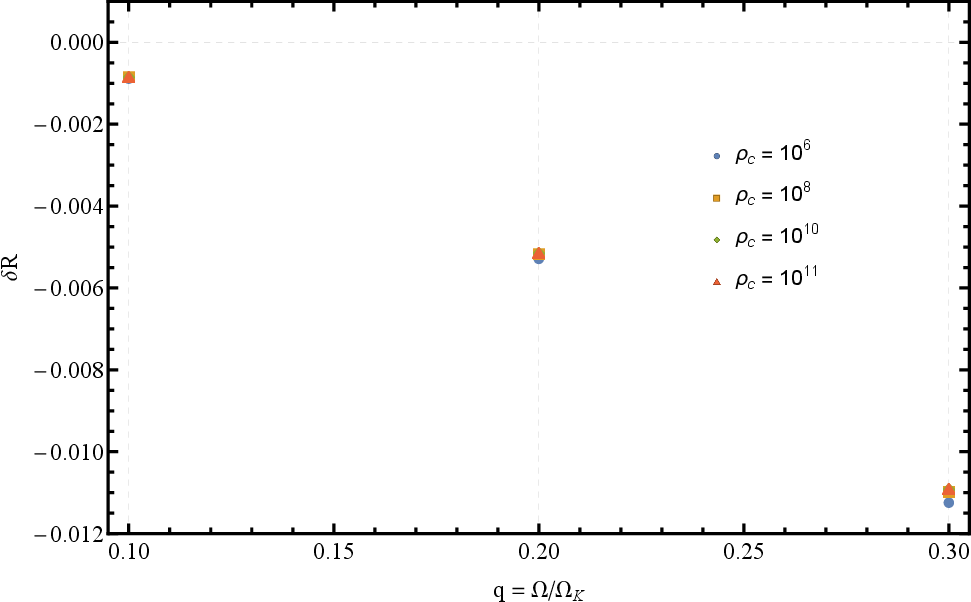}
\caption{Relative radius discrepancy between the one-dimensional reduced model and the quasi-two-dimensional reconstruction as a function of the rotation fraction $q=\Omega/\Omega_K$, for the sampled set of central densities $\rho_c=10^6,10^8,10^{10},10^{11}\,{\rm g\,cm^{-3}}$. Here the two-dimensional reference radius is the volume-equivalent radius $R_{\rm vol,2D}^{\rm approx}$. The discrepancy remains at the percent level over the tested range and grows monotonically with $q$.}
\label{fig:dR}
\end{figure}

The comparison is shown in Figs.~\ref{fig:dM} and \ref{fig:dR}, and the maximum
absolute discrepancies are summarized in the upper block of
Table~\ref{tab:quasi2D_combined}. Within the
tested slow-rotation range, the one-dimensional model remains close to the reconstructed
mass, while the radius shows the larger relative deviation. This result supports the use of
the reduction as a trend-level survey model, but it should not be interpreted as a validation
against a full two-dimensional equilibrium.

To separate the intrinsic rotational effect from the model--reconstruction discrepancy, we
also compute the fractional shifts of the one-dimensional rotating models relative to their
non-rotating baselines,
\begin{equation}
\Delta_M^{(0)}\equiv \frac{M(q)-M(0)}{M(0)},\qquad
\Delta_R^{(0)}\equiv \frac{R(q)-R(0)}{R(0)} .
\label{eq:deltaMR0_quasi2D}
\end{equation}
The maximum values over the sampled central-density set are listed in the lower block of
Table~\ref{tab:quasi2D_combined}. The radius shift is larger than the mass shift over
the tested range, reaching about $2.4\%$ at $q=0.3$, while the mass shift remains below
$1\%$.

\begin{table}[t]
\centering
\small
\setlength{\tabcolsep}{8pt}
\renewcommand{\arraystretch}{1.15}
\caption{Quasi-two-dimensional consistency diagnostics for the sampled rotation fractions
$q=\Omega/\Omega_K$. The upper block gives the maximum absolute discrepancies between
the one-dimensional reduced model and the quasi-two-dimensional reconstruction, while
the lower block gives the maximum absolute fractional shifts relative to the corresponding
non-rotating baselines.}
\label{tab:quasi2D_combined}
\begin{tabular}{ccc}
\midrule
\hline\hline
\multicolumn{3}{c}{Model--reconstruction discrepancies}
\\
\hline
$q$ & $\max|\delta M|$ & $\max|\delta R|$ \\
\midrule
0.1 & $8.73\times10^{-4}$ & $8.88\times10^{-4}$ \\
0.2 & $6.78\times10^{-4}$ & $5.29\times10^{-3}$ \\
0.3 & $3.03\times10^{-3}$ & $1.12\times10^{-2}$ \\
\hline
\multicolumn{3}{@{}c@{}}{Fractional shifts relative to the non-rotating baselines} \\
\midrule \hline
$q$ & $\max|\Delta_M^{(0)}|$ & $\max|\Delta_R^{(0)}|$ \\
\midrule
0.1 & $8.31\times10^{-4}$ & $2.47\times10^{-3}$ \\
0.2 & $3.35\times10^{-3}$ & $1.02\times10^{-2}$ \\
0.3 & $7.61\times10^{-3}$ & $2.39\times10^{-2}$ \\
\botrule
\end{tabular}
\end{table}

\FloatBarrier
\section{Static TOV scale assessment of the reduced rotational correction}
\label{sec:tov_scale_comparison}

The rotation--anisotropy mapping developed in this work is a Newtonian reduction. It
should not be interpreted as a relativistic rotating white-dwarf solution. Nevertheless, in
the high-density part of a white-dwarf sequence, static relativistic corrections can become
comparable in magnitude to rotational corrections. It is therefore useful to ask whether the
rotational shift produced by the present reduced model is negligible compared with the
leading static relativistic correction, or whether both effects can modify the same global
observables at the percent level.

For this purpose we compare the reduced Newtonian rotating sequences with a static,
non-rotating, isotropic TOV reference sequence. The
comparison is deliberately limited in scope. We do not combine rotation and relativity into
a single equilibrium model, and we do not claim to describe relativistic rotating white
dwarfs. Instead, the TOV sequence is used as a scale reference: it identifies the density
range in which the angle-averaged Newtonian rotational correction and the static
relativistic correction are of comparable size.

For the relativistic reference model we use the static spherically symmetric line element
\begin{equation}
ds^2 =
-e^{2\nu(r)}c^2dt^2
+
\left(1-\frac{2Gm(r)}{rc^2}\right)^{-1}dr^2
+
r^2d\Omega^2 .
\label{eq:tov_metric}
\end{equation}
The matter source is taken to be an isotropic perfect fluid. In order to keep the comparison
directly tied to the Newtonian sequences, we use the same Chandrasekhar pressure--density
relation as in the reduced model and approximate the relativistic energy density by
\begin{equation}
\varepsilon \simeq \rho c^2 .
\label{eq:tov_energy_density_approx}
\end{equation}
This approximation is sufficient for estimating the leading static relativistic correction to
the global mass and radius in the density range considered here. It should not, however, be
regarded as a complete thermodynamic model of a relativistic white dwarf, since internal
energy, composition-dependent corrections, finite-temperature effects, and inverse beta-decay
constraints are outside the scope of the present scale assessment.

With this convention, the TOV equations are written as
\begin{equation}
\frac{dm}{dr}=4\pi r^2\rho ,
\label{eq:tov_mass}
\end{equation}
\begin{equation}
\frac{dp}{dr}
=
-\frac{
G\left(\rho+p/c^2\right)
\left[m(r)+4\pi r^3p/c^2\right]
}
{
r^2\left[1-2Gm(r)/(rc^2)\right]
} .
\label{eq:tov_pressure}
\end{equation}
The surface of the relativistic reference configuration is defined by
\begin{equation}
p(R_{\rm TOV})=0,
\qquad
M_{\rm TOV}=m(R_{\rm TOV}) .
\label{eq:tov_surface}
\end{equation}
In this reference sequence both rotation and anisotropy are absent,
\begin{equation}
\Omega=0,\qquad \Delta=0 .
\label{eq:tov_reference_limit}
\end{equation}
Thus the TOV sequence is not an extension of the rotation--anisotropy mapping, but an
independent static reference computed with the same pressure--density input.

The purpose of this comparison is not to construct a relativistic rotating white-dwarf model, but to assess whether the rotational corrections retained by the reduced Newtonian framework remain significant relative to neglected relativistic effects.

The reduced rotational shifts are measured relative to the non-rotating Newtonian
baseline at the same central density:
\begin{equation}
\delta_M^{\rm rot}(f;\rho_c)
=
\frac{
M_{\rm Newt,rot}(f;\rho_c)-M_{\rm Newt,0}(\rho_c)
}
{
M_{\rm Newt,0}(\rho_c)
},
\qquad
\delta_R^{\rm rot}(f;\rho_c)
=
\frac{
R_{\rm Newt,rot}(f;\rho_c)-R_{\rm Newt,0}(\rho_c)
}
{
R_{\rm Newt,0}(\rho_c)
} .
\label{eq:rot_fractional_shifts}
\end{equation}
The corresponding static relativistic shifts are defined by
\begin{equation}
\delta_M^{\rm GR}(\rho_c)
=
\frac{
M_{\rm TOV,0}(\rho_c)-M_{\rm Newt,0}(\rho_c)
}
{
M_{\rm Newt,0}(\rho_c)
},
\qquad
\delta_R^{\rm GR}(\rho_c)
=
\frac{
R_{\rm TOV,0}(\rho_c)-R_{\rm Newt,0}(\rho_c)
}
{
R_{\rm Newt,0}(\rho_c)
} .
\label{eq:gr_fractional_shifts}
\end{equation}
For reference, we also compute the compactness and surface redshift of the static TOV
sequence,
\begin{equation}
C\equiv \frac{2GM}{Rc^2},
\qquad
z_s=
\left(1-\frac{2GM}{Rc^2}\right)^{-1/2}-1 .
\label{eq:compactness_redshift}
\end{equation}

Since the TOV reference sequence is static and isotropic, the present comparison should be interpreted as a scale assessment between distinct physical effects rather than as a combined relativistic-rotation calculation.

The main scale result is obtained at the high-density endpoint of the scan. For the
largest retained rotation proxy, \(f=0.35\), the reduced Newtonian model gives
\begin{equation}
\delta_M^{\rm rot}(f=0.35)\simeq 1.20\%,
\qquad
\delta_R^{\rm rot}(f=0.35)\simeq 3.23\% .
\label{eq:rot_shift_endpoint}
\end{equation}
At the same central density, the static TOV reference sequence gives
\begin{equation}
\delta_M^{\rm GR}\simeq -1.97\%,
\qquad
\delta_R^{\rm GR}\simeq -0.35\% .
\label{eq:gr_shift_endpoint}
\end{equation}
The two corrections therefore have different signs and different physical origins. The
reduced rotational term increases the radius and gives a positive mass shift in the
Newtonian sequence, while the static relativistic correction slightly compacts the
configuration and lowers the mass readout relative to the Newtonian baseline. Their
magnitudes, however, are not parametrically negligible relative to one another. This is the
main reason for including the TOV reference: in the compact, high-density part of the
sequence, Newtonian rotation and static relativistic gravity can both contribute at the
percent level to the same global observables.

The comparison therefore suggests that, in the high-density regime, neither rotational nor relativistic corrections can be assumed negligible a priori when percent-level accuracy in global observables is sought.

\begin{figure}
\centering
\includegraphics[width=0.6\textwidth]{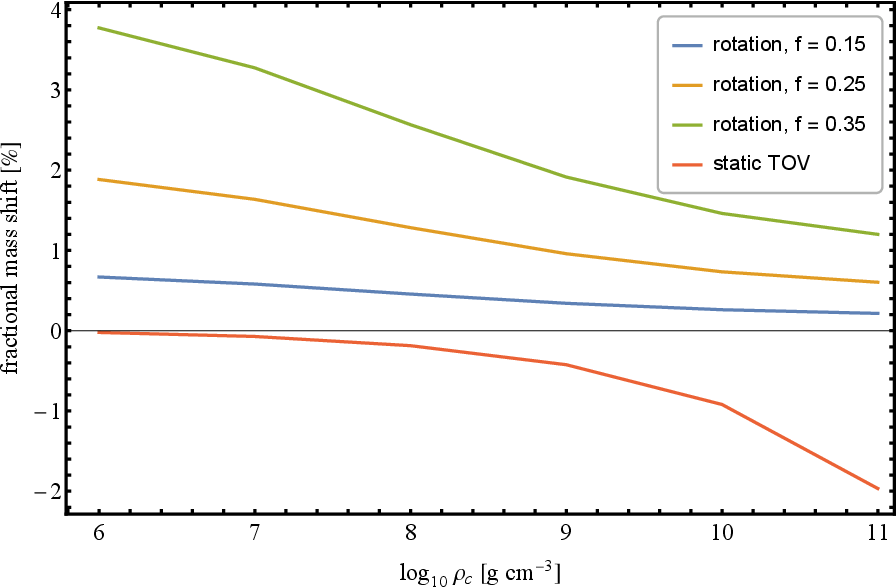}
\caption{Fractional mass shifts relative to the non-rotating Newtonian baseline as functions of
central density. The curves compare the reduced Newtonian rotating models for selected values
of \(f\) with the static isotropic TOV reference sequence.}
\label{fig:massShiftPlot}
\end{figure}

\begin{figure}
\centering
\includegraphics[width=0.6\textwidth]{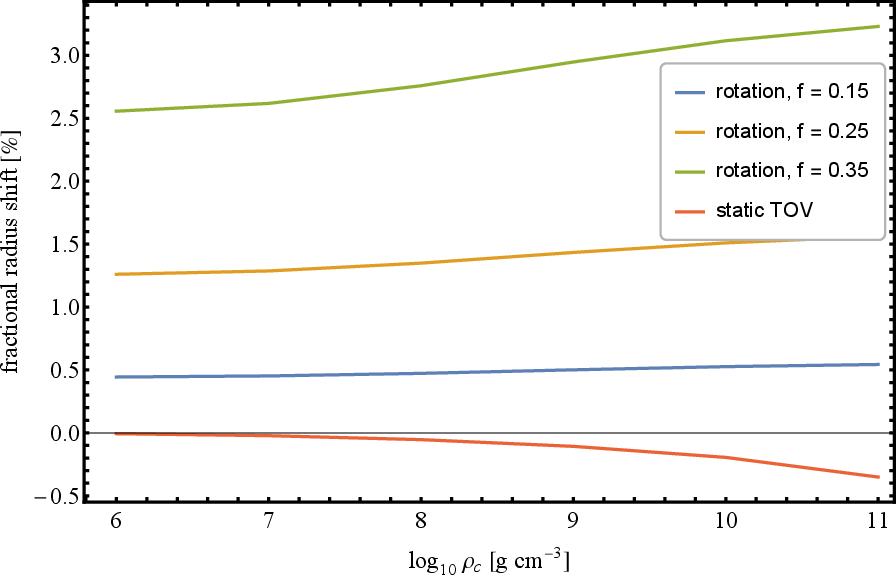}
\caption{Fractional radius shifts relative to the non-rotating Newtonian baseline as functions of
central density. The reduced rotational correction increases the radius, whereas the static TOV
correction gives a smaller negative shift over the high-density part of the sequence.}
\label{fig:radiusShiftPlot}
\end{figure}

Figures~\ref{fig:massShiftPlot} and \ref{fig:radiusShiftPlot} show the fractional mass and
radius shifts relative to the non-rotating Newtonian baseline. The mass channel displays a
competition between a positive rotational shift and a negative static relativistic shift. The
radius channel is more one-sided in the present parameter range: the reduced rotational
correction increases the radius, whereas the static TOV correction is smaller in magnitude
and negative over the high-density part of the sequence.

To make the relative importance of the two effects more explicit, we also plot
\begin{equation}
\log_{10}
\left|
\frac{\delta_M^{\rm rot}}{\delta_M^{\rm GR}}
\right|,
\qquad
\log_{10}
\left|
\frac{\delta_R^{\rm rot}}{\delta_R^{\rm GR}}
\right| .
\label{eq:ratio_diagnostics}
\end{equation}
Values above zero indicate that the reduced rotational correction is larger in magnitude
than the static TOV correction, while values below zero indicate the opposite. These ratio
diagnostics should not be interpreted as a hierarchy between rotation and relativity in a
complete stellar model. They only show, within the present controlled comparison, where the
two separately computed corrections become comparable.

The numerical readout used in this section is given in
Appendix~\ref{app:tov_readout}. The comparison supports a clear interpretation of the
reduced model. It is useful as a fast Newtonian benchmark because the rotational term is
explicit and its domain of applicability is monitored by internal diagnostics. At the same
time, when the rotational and static relativistic shifts both reach the percent level, precision
modelling should not treat either effect as automatically negligible. A quantitatively
complete treatment in that regime would require either a relativistic slow-rotation
calculation or fully rotating relativistic equilibria.

\begin{figure}
\centering
\includegraphics[width=0.6\textwidth]{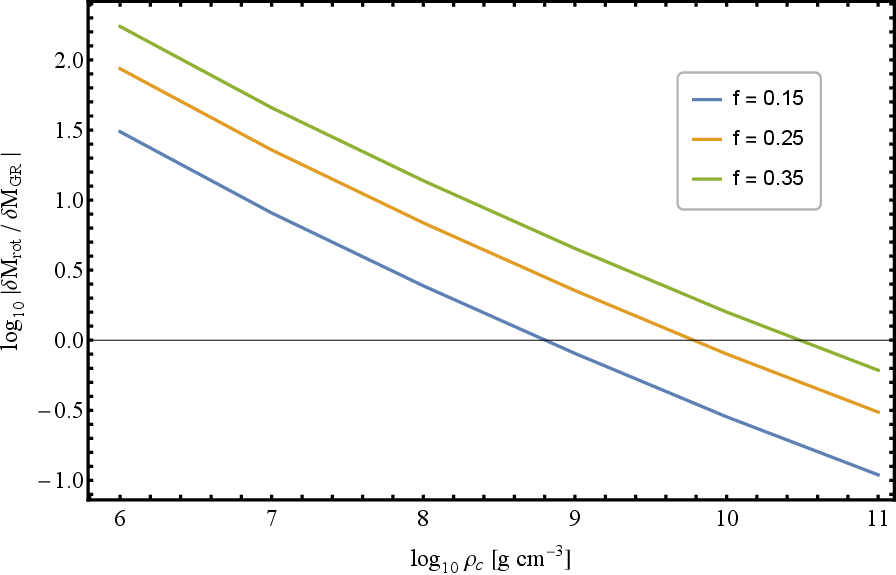}
\caption{Relative magnitude of the rotational and static relativistic mass shifts, shown as
\(\log_{10}|\delta_M^{\rm rot}/\delta_M^{\rm GR}|\). Values above zero indicate that the
reduced rotational correction is larger in magnitude than the static TOV correction.}
\label{fig:massDominancePlot}
\end{figure}

\begin{figure}
\centering
\includegraphics[width=0.6\textwidth]{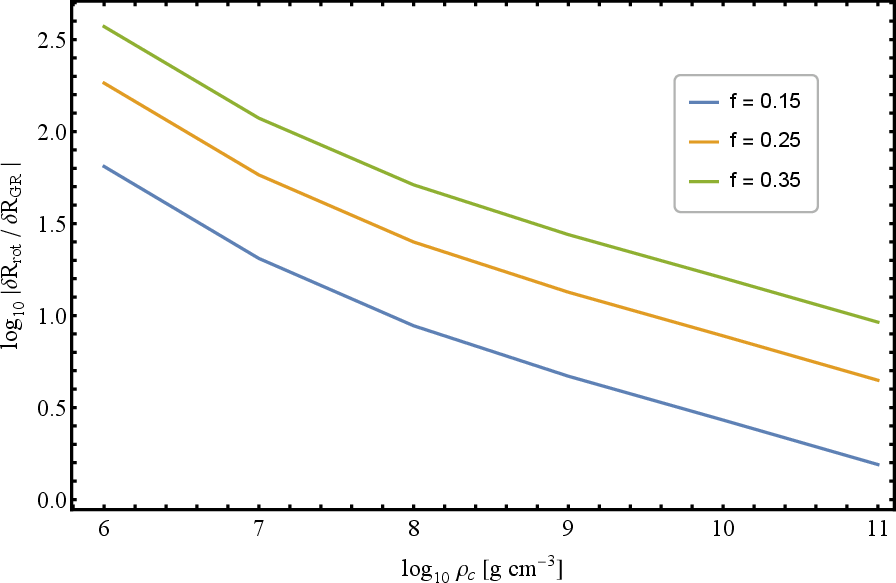}
\caption{Relative magnitude of the rotational and static relativistic radius shifts, shown as
\(\log_{10}|\delta_R^{\rm rot}/\delta_R^{\rm GR}|\). Values above zero indicate that the
reduced rotational correction is larger in magnitude than the static TOV correction.}
\label{fig:radiusDominancePlot}
\end{figure}

The convergence of the rotational and relativistic correction scales at the upper end of the explored density range provides a useful indicator of the validity limits of the reduced Newtonian framework. Specifically, it identifies the regime in which neglected relativistic contributions become comparable to the retained rotational corrections, thereby marking the onset of conditions under which a relativistic treatment is required for quantitatively accurate modeling.

\FloatBarrier

\section{Discussion and conclusions}
\label{sec:disc_conc}

We have constructed a one-dimensional Newtonian reduction for uniformly rotating cold
white dwarfs in which the angle-averaged centrifugal support is represented by an effective
anisotropic term,
\[
\Delta_{\rm rot}(r)=\frac{1}{3}\rho(r)\Omega^2 r^2 .
\]
This mapping keeps the spin frequency explicit and preserves a one-dimensional
hydrostatic system. Its interpretation is deliberately limited: $\Delta_{\rm rot}$ is not a
microscopic pressure anisotropy of the stellar matter, but an effective representation of the
spherically averaged centrifugal contribution.

The numerical sequences show the expected rotational response. As the rotation proxy
$f$ increases, the density and pressure profiles become slightly broader, and the global
families $M(\rho_c)$ and $R(\rho_c)$ move toward larger masses and radii relative to the
non-rotating Chandrasekhar sequence. In the high-density readout summarized in
Table~\ref{tab:summary}, both $M_{\rm lim}$ and $R_{\rm lim}$ increase monotonically with
$f$. For the largest retained value, $f=0.35$, the limiting-mass shift remains at the percent
level, while the corresponding configurations stay sub-Keplerian and the bulk-interior
measure $\mathcal{A}_{10^{-2}}$ remains below unity.

The limiting quantities should be understood in the operational sense used throughout the
paper. For the cold Newtonian Chandrasekhar equation of state, the mass approaches an
asymptotic value at large central density
\citep{Chandrasekhar1931,Chandrasekhar1939,ShapiroTeukolsky1983}. In the present
finite grid the discrete maxima occur at $\log_{10}\rho_c=11$. The quoted values of
$(M_{\rm lim},R_{\rm lim})$ are therefore high-density saturation estimates within the
adopted interval, not exact maximum-mass configurations and not a complete stability
boundary for rotating white dwarfs.

A central part of the construction is the applicability envelope. Since the scan parameter
$f=\Omega/\Omega_{K,0}$ is defined with respect to the non-rotating reference sequence, it is
not by itself a physical admissibility criterion. The relevant checks must be evaluated on
the rotating configurations. In the reported sequences, the sub-Keplerian condition is
safely satisfied; the more restrictive requirement is that the mapped anisotropic term remain
subdominant to the radial pressure in the bulk interior. This identifies the intended domain
of the model as slow-to-moderate rotation, where the reduced description can be used as a
controlled survey tool.

The quasi-two-dimensional reconstruction provides an auxiliary consistency check on this
interpretation. It does not solve the full axisymmetric Poisson--Euler problem, because the
background potential is kept spherical and the enthalpy distribution is reconstructed only
approximately. Within the sampled slow-rotation set, however, the one-dimensional masses
remain close to the reconstructed values at the $10^{-3}$ level, while the volume-equivalent
radii differ at the percent level. The increasing discrepancy with rotation is consistent with
the main limitation of the reduced model: it does not include the geometric deformation of
the star.

The static TOV scale assessment clarifies the relativistic status of the calculation. The
reduced rotational correction and the static relativistic correction arise from different
physical problems and should not be combined within the present Newtonian equations.
Nevertheless, when both are measured relative to the same non-rotating Newtonian
baseline, they can reach comparable percent-level magnitudes in the high-density part of
the sequence. This shows that, for compact white-dwarf models aiming at percent-level
accuracy, rotational and relativistic effects should be assessed together rather than assumed
to be hierarchically separated. In this sense, the present model is best viewed as a controlled
diagnostic benchmark for the rotational part of the problem, with the TOV comparison
indicating where a relativistic slow-rotation or fully rotating treatment becomes necessary.

The comparison with static TOV configurations further shows that relativistic and rotational corrections become comparable in the high-density regime, identifying the region where a purely Newtonian treatment approaches the limit of its quantitative applicability.

The main limitation of the model is geometric. The replacement
$\sin^2\theta\rightarrow\langle\sin^2\theta\rangle=2/3$ removes the latitude dependence of the
centrifugal force. As a result, the model cannot describe the equatorial bulge, polar
flattening, oblateness, or the true mass-shedding surface of a rotating white dwarf. The
rotating sector is also Newtonian and restricted to uniform rotation. Rapid rotation,
differential rotation, and relativistic rotational corrections require either self-consistent
axisymmetric Newtonian equilibria, relativistic slow-rotation schemes, or fully rotating
relativistic models.

These results place the reduced model in a well-defined regime of use. It is a Newtonian
benchmark with explicit applicability diagnostics, not a complete model of rotating white
dwarfs. Within this regime, it captures the leading spherical component of uniform rotation
and quantifies how this contribution shifts the mass and radius sequences. When the
rotational and relativistic corrections become comparable, the natural continuation is a
relativistic slow-rotation treatment or a self-consistent multidimensional equilibrium
calculation.
\FloatBarrier
\section*{Acknowledgments}
This research was funded by the Ministry of Science and Higher Education of the Republic of Kazakhstan (Zhas Galym 2025–2027, Grant No. $AP25795219$). The work of HQ was supported by UNAM-DGAPA-PAPIIT, grant No. 108225, and Conahcyt, grant No. CBF-2025-I-243.
\appendix

\section{Numerical readout for the static TOV scale assessment}
\label{app:tov_readout}

This appendix lists the numerical values used in the scale assessment discussed in
Sec.~\ref{sec:tov_scale_comparison}. All fractional shifts are measured relative to the
non-rotating Newtonian baseline at the same central density. Table~\ref{tab:tov_reference_readout}
gives the common Newtonian baseline and the static isotropic TOV reference sequence,
while Table~\ref{tab:rotation_shift_all_f} gives the rotation-induced fractional shifts for
the three rotation proxies used in the comparison.

\begin{table*}[t]
\centering
\small
\setlength{\tabcolsep}{5pt}
\renewcommand{\arraystretch}{1.15}
\caption{Non-rotating Newtonian baseline and static isotropic TOV reference sequence used in
the scale assessment. The fractional GR shifts are measured relative to the non-rotating
Newtonian model at the same central density.}
\label{tab:tov_reference_readout}
\begin{tabular}{cccccccc}
\hline\hline
$\log_{10}\rho_c$
&
$M_{\rm N,0}$ & $R_{\rm N,0}$
&
$M_{\rm TOV}$ & $R_{\rm TOV}$
&
$\delta M_{\rm GR}$ & $\delta R_{\rm GR}$ & $z_s$
\\
&
$[M_\odot]$ & $[{\rm km}]$
&
$[M_\odot]$ & $[{\rm km}]$
&
$[\%]$ & $[\%]$ &
\\
\hline
6  & 0.400799 & 10958.40 & 0.400712 & 10957.70 & -0.0217 & -0.0064 & $5.400\times10^{-5}$ \\
7  & 0.815116 & 7085.03  & 0.814530 & 7083.46  & -0.0718 & -0.0222 & $1.698\times10^{-4}$ \\
8  & 1.180470 & 4327.58  & 1.178270 & 4325.25  & -0.1866 & -0.0538 & $4.025\times10^{-4}$ \\
9  & 1.367190 & 2470.10  & 1.361390 & 2467.46  & -0.4238 & -0.1069 & $8.157\times10^{-4}$ \\
10 & 1.432700 & 1316.44  & 1.419530 & 1313.87  & -0.9192 & -0.1952 & $1.599\times10^{-3}$ \\
11 & 1.450720 & 662.77   & 1.422200 & 660.45   & -1.9661 & -0.3500 & $3.195\times10^{-3}$ \\
\hline\hline
\end{tabular}
\end{table*}

\begin{table*}[t]
\centering
\small
\setlength{\tabcolsep}{6pt}
\renewcommand{\arraystretch}{1.15}
\caption{Fractional shifts produced by the reduced rotational correction. The shifts are
computed relative to the non-rotating Newtonian baseline at the same central density.}
\label{tab:rotation_shift_all_f}
\begin{tabular}{ccccccc}
\hline\hline
$\log_{10}\rho_c$
&
\multicolumn{2}{c}{$f=0.15$}
&
\multicolumn{2}{c}{$f=0.25$}
&
\multicolumn{2}{c}{$f=0.35$}
\\
\cline{2-7}
&
$\delta M_{\rm rot}$ & $\delta R_{\rm rot}$
&
$\delta M_{\rm rot}$ & $\delta R_{\rm rot}$
&
$\delta M_{\rm rot}$ & $\delta R_{\rm rot}$
\\
&
$[\%]$ & $[\%]$
&
$[\%]$ & $[\%]$
&
$[\%]$ & $[\%]$
\\
\hline
6  & 0.6689 & 0.4440 & 1.1963 & 0.7967 & 3.7722 & 2.5561 \\
7  & 0.5813 & 0.4524 & 1.0394 & 0.8125 & 3.2746 & 2.6180 \\
8  & 0.4564 & 0.4729 & 0.8158 & 0.8503 & 2.5664 & 2.7584 \\
9  & 0.3412 & 0.5010 & 0.6097 & 0.9022 & 1.9143 & 2.9471 \\
10 & 0.2613 & 0.5264 & 0.4666 & 0.9488 & 1.4610 & 3.1159 \\
11 & 0.2153 & 0.5436 & 0.3843 & 0.9803 & 1.2002 & 3.2296 \\
\hline\hline
\end{tabular}
\end{table*}

The values in Tables~\ref{tab:tov_reference_readout} and
\ref{tab:rotation_shift_all_f} underlie the fractional-shift and ratio diagnostics shown in
Figs.~\ref{fig:massShiftPlot}--\ref{fig:radiusDominancePlot}. The purpose of the appendix is
only to provide the numerical readout used in those plots; the physical interpretation of the
comparison is given in Sec.~\ref{sec:tov_scale_comparison}.

\clearpage

\bibliographystyle{unsrt}
\bibliography{References_with_TOV}

@article{Chandrasekhar1931,
  author  = {Chandrasekhar, S.},
  title   = {The Maximum Mass of Ideal White Dwarfs},
  journal = {The Astrophysical Journal},
  volume  = {74},
  pages   = {81--82},
  year    = {1931},
  doi     = {10.1086/143324}
}

@book{Chandrasekhar1939,
  author    = {Chandrasekhar, S.},
  title     = {An Introduction to the Study of Stellar Structure},
  publisher = {University of Chicago Press},
  address   = {Chicago},
  year      = {1939}
}

@article{HamadaSalpeter1961,
  author  = {Hamada, T. and Salpeter, E. E.},
  title   = {Models for Zero-Temperature Stars},
  journal = {The Astrophysical Journal},
  volume  = {134},
  pages   = {683--698},
  year    = {1961},
  doi     = {10.1086/147195}
}

@article{Nauenberg1972,
  author  = {Nauenberg, Michael},
  title   = {Analytic Approximations to the Mass-Radius Relation and Energy of Zero-Temperature Stars},
  journal = {The Astrophysical Journal},
  volume  = {175},
  pages   = {417--430},
  year    = {1972},
  doi     = {10.1086/151568}
}

@book{ShapiroTeukolsky1983,
  author    = {Shapiro, Stuart L. and Teukolsky, Saul A.},
  title     = {Black Holes, White Dwarfs, and Neutron Stars: The Physics of Compact Objects},
  publisher = {Wiley-Interscience},
  address   = {New York},
  year      = {1983}
}

@book{Warner1995CVs,
  author    = {Warner, Brian},
  title     = {Cataclysmic Variable Stars},
  series    = {Cambridge Astrophysics Series},
  volume    = {28},
  publisher = {Cambridge University Press},
  address   = {Cambridge},
  year      = {1995},
  doi       = {10.1017/CBO9780511586491}
}

@article{Kawaler2015,
  author  = {Kawaler, Steven D.},
  title   = {Rotation of White Dwarf Stars},
  journal = {Astronomical Society of the Pacific Conference Series},
  volume  = {493},
  pages   = {65},
  year    = {2015}
}

@article{Hermes2017ApJS,
  author  = {Hermes, J. J. and Gänsicke, B. T. and Kawaler, S. D. and Greiss, S. and Tremblay, P.-E. and Gentile Fusillo, N. P. and Raddi, R. and Fanale, S. M. and Bell, K. J. and Dennihy, E. and Fuchs, J. T. and Dunlap, B. H. and Clemens, J. C. and Montgomery, M. H. and Winget, D. E. and Chote, P. and Marsh, T. R. and Redfield, S.},
  title   = {White Dwarf Rotation as a Function of Mass and a Dichotomy of Mode Line Widths: Kepler Observations of 27 Pulsating DA White Dwarfs through K2 Campaign 8},
  journal = {The Astrophysical Journal Supplement Series},
  volume  = {232},
  number  = {2},
  pages   = {23},
  year    = {2017},
  doi     = {10.3847/1538-4365/aa8bb5},
  eprint  = {1709.07004},
  archivePrefix = {arXiv},
  primaryClass  = {astro-ph.SR}
}

@article{Brinkworth2013ApJ,
  author  = {Brinkworth, Carolyn S. and Burleigh, M. R. and Lawrie, Katherine and Marsh, T. R. and Knigge, Christian},
  title   = {Measuring the Rotation Rates of Isolated Magnetic White Dwarfs},
  journal = {The Astrophysical Journal},
  volume  = {773},
  number  = {1},
  pages   = {47},
  year    = {2013},
  doi     = {10.1088/0004-637X/773/1/47}
}

@article{Ferrario2015SSRv,
  author  = {Ferrario, Lilia and de Martino, Domitilla and Gänsicke, Boris T.},
  title   = {Magnetic White Dwarfs},
  journal = {Space Science Reviews},
  volume  = {191},
  pages   = {111--169},
  year    = {2015},
  doi     = {10.1007/s11214-015-0152-0},
  eprint  = {1504.08072},
  archivePrefix = {arXiv},
  primaryClass  = {astro-ph.SR}
}

@article{Kilic2021ApJL,
  author  = {Kilic, Mukremin and Kosakowski, Alekzander and Moss, Adam G. and Bergeron, P. and Conly, Annamarie A.},
  title   = {An Isolated White Dwarf with a 70 s Spin Period},
  journal = {The Astrophysical Journal Letters},
  volume  = {923},
  number  = {1},
  pages   = {L6},
  year    = {2021},
  doi     = {10.3847/2041-8213/ac3b60}
}

@book{Chandrasekhar1969,
  author    = {Chandrasekhar, S.},
  title     = {Ellipsoidal Figures of Equilibrium},
  publisher = {Yale University Press},
  address   = {New Haven},
  year      = {1969}
}

@book{Tassoul1978,
  author    = {Tassoul, Jean-Louis},
  title     = {Theory of Rotating Stars},
  publisher = {Princeton University Press},
  address   = {Princeton, NJ},
  year      = {1978}
}

@article{OstrikerBodenheimer1968,
  author  = {Ostriker, Jeremiah P. and Bodenheimer, Peter},
  title   = {Rapidly Rotating Stars. II. Massive White Dwarfs},
  journal = {The Astrophysical Journal},
  volume  = {151},
  pages   = {1089--1098},
  year    = {1968},
  doi     = {10.1086/149494}
}

@article{Hachisu1986ApJS,
  author  = {Hachisu, Izumi},
  title   = {A Versatile Method for Obtaining Structures of Rapidly Rotating Stars},
  journal = {The Astrophysical Journal Supplement Series},
  volume  = {61},
  pages   = {479--507},
  year    = {1986},
  doi     = {10.1086/191118}
}

@article{EriguchiMuller1985,
  author  = {Eriguchi, Y. and Müller, E.},
  title   = {A General Computational Method for Obtaining Equilibria of Self-Gravitating and Rotating Gaseous Bodies},
  journal = {Astronomy and Astrophysics},
  volume  = {146},
  pages   = {260--268},
  year    = {1985}
}

@article{KomatsuEriguchiHachisu1989,
  author  = {Komatsu, Hideyuki and Eriguchi, Yoshiharu and Hachisu, Izumi},
  title   = {Rapidly Rotating General Relativistic Stars. I. Numerical Method and Its Application to Uniformly Rotating Polytropes},
  journal = {Monthly Notices of the Royal Astronomical Society},
  volume  = {237},
  pages   = {355--379},
  year    = {1989},
  doi     = {10.1093/mnras/237.2.355}
}

@article{Hartle1967,
  author  = {Hartle, James B.},
  title   = {Slowly Rotating Relativistic Stars. I. Equations of Structure},
  journal = {The Astrophysical Journal},
  volume  = {150},
  pages   = {1005--1029},
  year    = {1967},
  doi     = {10.1086/149400}
}

@article{HartleThorne1968,
  author  = {Hartle, James B. and Thorne, Kip S.},
  title   = {Slowly Rotating Relativistic Stars. II. Models for Neutron Stars and Supermassive Stars},
  journal = {The Astrophysical Journal},
  volume  = {153},
  pages   = {807--834},
  year    = {1968},
  doi     = {10.1086/149707}
}

@article{CookShapiroTeukolsky1992,
  author  = {Cook, Gregory B. and Shapiro, Stuart L. and Teukolsky, Saul A.},
  title   = {Rapidly rotating neutron stars in general relativity: Realistic equations of state},
  journal = {The Astrophysical Journal},
  volume  = {424},
  pages   = {823--845},
  year    = {1994},
  doi     = {10.1086/173934}
}

@article{Stergioulas2003,
  author  = {Stergioulas, Nikolaos},
  title   = {Rotating Stars in Relativity},
  journal = {Living Reviews in Relativity},
  volume  = {6},
  pages   = {3},
  year    = {2003},
  doi     = {10.12942/lrr-2003-3}
}

@book{FriedmanStergioulas2013,
  author    = {Friedman, John L. and Stergioulas, Nikolaos},
  title     = {Rotating Relativistic Stars},
  publisher = {Cambridge University Press},
  address   = {Cambridge},
  year      = {2013},
  doi       = {10.1017/CBO9780511977596}
}

@article{Boshkayev2013ApJ,
  author  = {Boshkayev, Kuantay and Rueda, Jorge A. and Ruffini, Remo and Siutsou, Ivan},
  title   = {On General Relativistic Uniformly Rotating White Dwarfs},
  journal = {The Astrophysical Journal},
  volume  = {762},
  number  = {2},
  pages   = {117},
  year    = {2013},
  doi     = {10.1088/0004-637X/762/2/117},
  eprint  = {1204.2070},
  archivePrefix = {arXiv},
  primaryClass  = {astro-ph.SR}
}

@article{Rotondo2011,
  author  = {Rotondo, M. and Rueda, Jorge A. and Ruffini, Remo and Xue, S.-S.},
  title   = {Relativistic Feynman-Metropolis-Teller Theory for White Dwarfs in General Relativity},
  journal = {Physical Review D},
  volume  = {84},
  pages   = {084007},
  year    = {2011},
  doi     = {10.1103/PhysRevD.84.084007},
  eprint  = {1012.0154},
  archivePrefix = {arXiv},
  primaryClass  = {astro-ph.SR}
}

@article{Boshkayev2014,
  author  = {Boshkayev, Kuantay and Rueda, Jorge A. and Ruffini, Remo and Siutsou, Ivan},
  title   = {General Relativistic White Dwarfs and Their Astrophysical Applications},
  journal = {Journal of the Korean Physical Society},
  volume  = {65},
  pages   = {855--862},
  year    = {2014},
  doi     = {10.3938/jkps.65.855},
  eprint  = {1412.8208},
  archivePrefix = {arXiv},
  primaryClass  = {astro-ph.SR}
}

@article{BowersLiang1974,
  author  = {Bowers, Richard L. and Liang, E. P. T.},
  title   = {Anisotropic Spheres in General Relativity},
  journal = {The Astrophysical Journal},
  volume  = {188},
  pages   = {657--665},
  year    = {1974},
  doi     = {10.1086/152760}
}

@article{HerreraSantos1997,
  author  = {Herrera, L. and Santos, N. O.},
  title   = {Local Anisotropy in Self-Gravitating Systems},
  journal = {Physics Reports},
  volume  = {286},
  number  = {2},
  pages   = {53--130},
  year    = {1997},
  doi     = {10.1016/S0370-1573(96)00042-7}
}

@article{DevGleiser2002,
  author  = {Dev, Krsna and Gleiser, Marcelo},
  title   = {Anisotropic Stars: Exact Solutions},
  journal = {General Relativity and Gravitation},
  volume  = {34},
  pages   = {1793--1818},
  year    = {2002},
  doi     = {10.1023/A:1020707906543}
}

@article{Tolman1939,
  author  = {Tolman, Richard C.},
  title   = {Static Solutions of Einstein's Field Equations for Spheres of Fluid},
  journal = {Physical Review},
  volume  = {55},
  number  = {4},
  pages   = {364--373},
  year    = {1939},
  doi     = {10.1103/PhysRev.55.364}
}

@article{OppenheimerVolkoff1939,
  author  = {Oppenheimer, J. R. and Volkoff, G. M.},
  title   = {On Massive Neutron Cores},
  journal = {Physical Review},
  volume  = {55},
  number  = {4},
  pages   = {374--381},
  year    = {1939},
  doi     = {10.1103/PhysRev.55.374}
}

@article{LaiRasioShapiro1993,
  author  = {Lai, Dong and Rasio, Frederic A. and Shapiro, Stuart L.},
  title   = {Ellipsoidal Figures of Equilibrium: Compressible Models},
  journal = {The Astrophysical Journal Supplement Series},
  volume  = {88},
  pages   = {205--252},
  year    = {1993},
  doi     = {10.1086/191823}
}

@article{OstrikerPeebles1973,
  author  = {Ostriker, Jeremiah P. and Peebles, P. J. E.},
  title   = {A Numerical Study of the Stability of Flattened Galaxies: or, Can Cold Galaxies Survive?},
  journal = {The Astrophysical Journal},
  volume  = {186},
  pages   = {467--480},
  year    = {1973},
  doi     = {10.1086/152513}
}

\end{document}